\documentclass[AMS,STIX1COL]{WileyNJD-v2}
\usepackage{moreverb}

\usepackage{graphicx}
\usepackage{dcolumn}
\usepackage{bm}
\usepackage{amsmath}
\usepackage{color}
\usepackage{longtable}

\newcommand\BibTeX{{\rmfamily B\kern-.05em \textsc{i\kern-.025em b}\kern-.08em
T\kern-.1667em\lower.7ex\hbox{E}\kern-.125emX}}

\articletype{Article Type}%

\received{<day> <Month>, <year>}
\revised{<day> <Month>, <year>}
\accepted{<day> <Month>, <year>}

\newcommand{\Ra}{\text{Ra}}   
\newcommand{\Rey}{\text{Re}}  
\newcommand{\Pran}{\text{Pr}} 
\newcommand{\Ro}{\text{Ro}}   

\newcommand{\red}[1]{{\textcolor{black}{#1}}}
\newcommand{\blue}[1]{{\textcolor{black}{#1}}}

\newcommand{\beq}{\begin{equation}}
\newcommand{\eeq}{\end{equation}}

\begin{document}

\title{Data-driven identification of the spatio-temporal structure of turbulent flows by streaming Dynamic Mode Decomposition}

\author[1,2]{Rui Yang}

\author[1]{Xuan Zhang}

\author[1]{Philipp Reiter}

\author[1,2]{Detlef Lohse}

\author[1]{Olga Shishkina}

\author[3]{Moritz Linkmann*}

\authormark{R. Yang \textsc{et al}}

\address[1]{\orgname{Max Planck Institute for Dynamics and Self-Organisation, Am Fassberg 17, 37077 G\"ottingen, Germany}}

\address[2]{\orgname{Physics of Fluids Group, Max Planck Center for Complex Fluid Dynamics, MESA+ Institute and J.M.~Burgers Center for Fluid Dynamics, University of Twente, P.O. Box 217, 7500 AE Enschede, The Netherlands}}

\address[3]{ \orgname{School of Mathematics and Maxwell Institute for Mathematical Sciences, University of Edinburgh, Edinburgh, EH9 3FD, United Kingdom}}

\corres{*Moritz Linkmann, School of Mathematics and Maxwell Institute for Mathematical Sciences, University of Edinburgh, Edinburgh, EH9 3FD, United Kingdom \email{Moritz.Linkmann@ed.ac.uk}}

\abstract[Abstract]{Streaming Dynamic Mode Decomposition (sDMD) (Hemati {\it et al.}, Phys.~Fluids 
{\bf 26}
(2014)) is a low-storage version of Dynamic Mode Decomposition (DMD) (Schmid, J.~Fluid Mech. {\bf 656} 
(2010)), a data-driven method to extract spatio-temporal flow patterns. 
Streaming DMD avoids storing the entire data sequence in memory by approximating the dynamic modes through incremental updates with new available data.  
In this paper, we use sDMD to identify and extract dominant spatio-temporal structures of different turbulent flows, requiring the analysis of large datasets.
First, the efficiency and accuracy of sDMD are compared to the classical DMD, using a publicly available test dataset that consists of velocity field
snapshots obtained by direct numerical simulation of a wake flow behind a cylinder.  
Streaming DMD not only reliably reproduces the most important dynamical features of the flow; our calculations also highlight its advantage in terms of the required computational resources.
We subsequently use sDMD to analyse three different turbulent flows that all show some degree of large-scale coherence: 
rapidly rotating Rayleigh--B\'enard convection, horizontal convection and the asymptotic suction boundary layer. 
Structures of different frequencies and spatial extent can be clearly separated, and the prominent features of the dynamics are captured with just a few dynamic modes.
In summary, we demonstrate that sDMD is a powerful tool for the identification of spatio-temporal structures in a wide range of turbulent flows.}



\maketitle

\section{Introduction}

Coherent structures at different spatial and temporal scales are a prominent
feature of many turbulent fluid flows occurring in nature and in engineering
applications \cite{Yaglom1967,Holmes1996,FazleHussain1986}. Examples include
large-scale vortices, wakes, convection rolls and thermal plumes in
Rayleigh--B\'enard convection (RBC) \cite{ahlers2009,Lohse2010}, Taylor rolls
in Taylor--Couette flow \cite{grossmann2016high}, jets, travelling waves,
very-large scale motions \cite{Hutchins2007a,Smits2011} and low-momentum zones
\cite{Meinhart1995} that develop in wall-bounded turbulent boundary layers
(BLs). These structures are known to have manifold significant effects in
turbulent flows, for instance influence on heat and mass transport, the
occurrence of extreme fluctuations or enhanced drag due their to interaction
with near-wall dynamics in turbulent BLs \cite{Monty2007, Marusic2010a,
Katul2019}.  Improving our knowledge of multi-scale spatio-temporal coherence
and the underlying physics is of paramount importance as it would lead to a
better fundamental understanding of turbulence, specifically in terms of
model-building and turbulence control. However, the co-existence of several
coherent structures makes the identification and the extraction of particular
spatio-temporal features difficult, which led to a growing need for data-driven
methods designed to identify and extract patterns.

Modal decomposition, as an umbrella term for a variety of structurally similar
methods, identifies structures by decomposing a given dataset in a suitable set
of basis functions, or modes.  Fourier analysis constitutes perhaps the most
well-known and widely used example of a modal decomposition technique. A more
sophisticated example is Proper Orthogonal Decomposition (POD)
\cite{sirovich1987pod, berkooz1993proper,podvin1998low,Rowley2004,bailon2012low}, where each
mode describes a flow structure according to its energy content. However, as
the POD modes do not produce a separated and compact
signal in frequency space, they
usually contain more than one characteristic frequency and thus cannot yield
information on frequential coherence.

Dynamic Mode Decomposition (DMD), by contrast, decomposes a dataset into
spatio-temporal coherent structures \cite{Schmid2010} with {\emph{dynamic modes} }
obtained as eigenmodes of a high-dimensional linear approximation of the
dynamics. More precisely, DMD has solid mathematical foundations in the context
of nonlinear dynamical systems theory. Under certain conditions it represents a
finite-dimensional approximation of the Koopman operator
\cite{Rowley2009,Williams2015}, a linear but infinite-dimensional
representation of a nonlinear dynamical system
\cite{Koopman1931,Mezic2005,Mezic2019}. 
DMD results have an intuitive physical
interpretation as each dynamic mode corresponds to a single frequency and
growth or decay rate. Therefore, it is a well-suited data-driven method for the
analysis of complex datasets and model reduction. Since its introduction by
Peter Schmid in 2010 \cite{Schmid2010}, DMD has had a history of successful
applications in fluid dynamics such as obtaining low-dimensional dynamic model
of the cylinder wake flow \cite{Tissot2014,bagheri2013}, generating good
initial guesses for unstable periodic orbits in turbulent channel flow
\cite{Page2020}, flow control \cite{Brunton2015,Proctor2016,Rowley2017},
aerodynamics \cite{Ghoreyshi2014}, and more general in pattern recognition
\cite{Jovanovic2014,Brunton2016}. An overview of the development of DMD extensions and applications thereof in the context of fluid dynamics is given in an upcoming review article by Peter Schmid \cite{schmid2022annurev}.

Most DMD applications consist of post-processing a time series of experimental
or computational data, where most implementations process the entire data sequence at once. 
However, the size of highly resolved
turbulent flow data usually precludes saving or loading the entire dataset into
memory. Therefore, only a few studies so far have applied DMD to highly
turbulent flows. These constraints can be circumvented by a DMD implementation
that allows for incremental data updates
\cite{Hemati2014,Anantharamu2019,Zhang2019}, such that the DMD calculation
proceeds alongside the main \red{data acquisition} process such as Direct Numerical Simulations (DNS) or
real-time Particle Image Velocimetry (PIV).  {\emph{Streaming} } DMD (sDMD)
\cite{Hemati2014} is such a method, which requires only two data samples at a
given instant in time and converges to the same results as classical DMD. In
what follows we focus on sDMD as a promising method for the analysis of
turbulent flows. 

The present article is intended to serve \red{three} 
purposes:
    (a) to demonstrate the applicability of streaming DMD \red{across different} large datasets of highly turbulent flows relevant to fundamental science and engineering applications,
    (b) to analyse the \red{large-scale} spatio-temporal dynamics of the flow in sub-domains of particular interest, 
    \red{(c) to demonstrate the robustness of the algorithm with respect to different degrees of downsampling, which allows for analyses of very large datasets to be carried out efficiently on local desktop machines.} 

The streaming version of the DMD algorithm \cite{Hemati2014} is applied to three datasets consisting of time series obtained in DNS of three different turbulent flows: 
rapidly rotating RBC, horizontal convection (HC)
and the asymptotic suction boundary layer (ASBL). Despite their physical differences, these three system share a few features that render them interesting and suitable as test cases. We demonstrate the advantages of sDMD for the analysis of turbulent flows, \red{with a particular focus on large-scale spatio-temporal data features}.

First, all three cases are paradigmatic examples of fluid-dynamic systems of
interest in geophysical fluid dynamics and engineering
applications. Rapidly rotating RBC is of relevance whenever
rotation and thermal convection are the key physical processes
\cite{ahlers2009, siggia1994}, such as in the dynamics of
planetary cores. Horizontal convection
\cite{Spiegel1971,Scott2001,Hughes2008,shishkina2016heat} occurs in the ocean which
is mostly heated and cooled by its upper surface being in
contact with the atmosphere.  The ASBL
\cite{Jones1963,Schlichting1979} is a flat-plate BL with a
constant BL thickness in the streamwise direction. The latter
is achieved by removing fluid through the pores in the bottom
plate, \red{a well-known technique for BL stabilisation.}  
\red{Furthermore, due to the constant BL thickness t}he ASBL allows the 
application of techniques developed for parallel wall-bounded shear flows to an open
flow.

Second, all three systems host spatio-temporally coherent
structures. In rapidly rotating RBC, this is the boundary zonal flow, a
large-scale travelling wave structure confined to the lateral near-wall region \cite{Zhang2020,favier2020robust, Shishkina2020}. 
HC features two characteristic processes that operate on very different time scales, i.~e., plume emission and slow oscillatory dynamics in the bulk \cite{Reiter2020}, with the former one being an order of magnitude faster than the latter one. The ASBL shows  coherent low momentum zones in the free stream, as do many wall bounded shear flows and freely evolving BLs \cite{Meinhart1995}, in the present dataset with a slow spanwise drift.

Third, the size of the datasets presents challenges in all three cases that can be mitigated by the incremental nature of \blue{sDMD}. For rapidly rotating RBC and HC, the fine grids required to properly resolve the small-scale turbulent dynamics result in large datasets, as usual for DNS of turbulent flows. In case of ASBL, a further difficulty lies in the slow dynamics of the low-momentum zone, as an analysis thereof requires very long time series. 

This article is organised as follows. Section \ref{sec:dmd} provides a summary
of both, classical DMD \cite{Schmid2010} and streaming DMD \cite{Hemati2014},
where we highlight few subtle differences concerning technical steps and
compare our implementations of DMD and sDMD using a standard publicly available
dataset -- DNS of a developing von-K\'arm\'an vortex street. The main results
of our analysis concerning turbulent flows are contained in
Sec.~\ref{sec:results}, beginning with rapidly rotating Rayleigh--B\'enard
convection in Sec.~\ref{sec:rbc}, followed by horizontal convection in
Sec.~\ref{sec:hc}  and the asymptotic suction BL in Sec.~\ref{sec:asbl}. 
The paper ends with conclusions and an outlook in Sec.~\ref{sec:conclusions}.

\section{Dynamic mode decomposition} 
\label{sec:dmd}
Before describing the specific features and advantages of streaming DMD (sDMD) \cite{Hemati2014}, we briefly summarise
the basic ideas and the classical singular value decomposition (SVD) based DMD algorithm \cite{Schmid2010}. 
For simplicity we restrict ourselves here to the case of equidistant
data sequences, for a more general discussion see \cite{kutz2016dynamic}.
Consider a time series of spatially resolved measurement results recorded at a fixed sampling rate $1/\Delta t$ resulting in, say, 
$N$ equidistant snapshots. Let us further assume that the possibly multidimensional data in each snapshot is flattened into a corresponding $M$-dimensional real vector, such that the time series can be represented by an ordered sequence  $(\bm{x}_k)_{\{k=1, \hdots, N\}}$ 
of column vectors $\bm{x}_k \in \mathbb{R}^M$ for $k \in \{1, \hdots, N\}$.
In the present context $\bm{x}_k$ would represent the $k^{\rm th}$ velocity field in a series of $N$ measurements, 
hence in particular for highly resolved three-dimensional flow fields \blue{ $M = 3 N_p^3$, where $N_p$ is the number of grid points,}
can quickly become very large. We will come back to this point in due course.

The assumption DMD relies upon is the existence of a linear operator $\bm{A} \in \mathbb{R}^{M \times M}$ which approximates the nonlinear dynamics across the interval $\Delta t$, that is
\beq
\label{eq:arnoldi}
\bm{x}_{k + 1} = \bm{A} \bm{x}_{k} + \bm{\varepsilon}_k \quad \text{for all} \quad k \in \{1, \hdots, N-1\} \ .
\eeq
Here, crucially, $\bm{A}$ does not depend on $k$. Finally, $\bm{\varepsilon}_k$ denotes an error term that is assumed to be small. 
The validity of this assumption
depends \blue{to some extent} on the ratio of the characteristic time scale of the observed nonlinear dynamics and the sampling interval $\Delta t$, \blue{but most importantly
on the potential to describe the dynamics by a linear surrogate model (i.e., the degree of nonlinearity)}. In practice, $\bm{A}$ is chosen by regression over the available data by least-squares minimisation of the $\bm{\varepsilon}_k$ \cite{kutz2016dynamic}.  
Since the operator $\bm{A}$ describes the spatio-temporal dynamics of the system, its eigenvectors, known as {\emph{dynamic modes} } 
or somewhat tautologically DMD modes, may be used to disentangle complex spatio-temporal
dynamics and to construct low-dimensional models. In what follows we summarize how the dynamic modes may be determined from the 
data sequence $(\bm{x}_k)_{\{k=1, \hdots, N\}}$, following an SVD-based approach as this is what is mostly used in practice 
owing to numerical stability concerns with the more fundamental Krylov-subspace-type approach and for reasons of 
computational cost reduction. Further details can be found in the original work by \cite{Schmid2010} and the 
textbook by \cite{kutz2016dynamic}. 

\subsection{SVD-based DMD}
\label{sec:svd-dmd}
For what follows it is convenient to combine data sequences that consist of
$N-1$ samples and are shifted forwards in time by $\Delta t$, that is
$(\bm{x}_k)_{\{k=1, \hdots, N-1\}}$ and $(\bm{x}_k)_{\{k=2, \hdots, N\}}$, into
$M \times (N-1)$-dimensional matrices
\begin{align}
	\label{eq:X1}
	\bm{X} = \bm{X}_1^{N-1} = (x_{jk}) & : = \left( \bm{x}_1 \ \bm{x}_2 \ \cdots \ \bm{x}_{N-1} \right) \ , \\
	\label{eq:X2}
	\bm{Y} = \bm{X}_2^{N}   = (y_{jk}) & : = \left( \bm{x}_2 \ \bm{x}_3 \ \cdots \ \bm{x}_{N} \right) \ ,
\end{align}
where $j \in {1, \hdots,  M}$ is the spatial index and $k$ the temporal index.
Then Eq.~\eqref{eq:arnoldi} implies 
\beq
\label{eq:data-evol}
\bm{Y} = \bm{A} \bm{X} + \bm{R} \ ,
\eeq
where 
$\bm{R} =
(\varepsilon_{jk})$ is the matrix of residuals.  The best-fit solution for
$\bm{A}$ with respect to least-squares minimization of $\bm{R}$ is given by
$\bm{A} = \bm{Y}\bm{X}^+$, where $\bm{X}^+$ is
the pseudo-inverse of $\bm{X}$. 
In practice, and in particular in fluid dynamics, $M \gg N$ as the dimension $M$ of the spatial samples usually exceeds
the number of temporal samples $N$ by far. Hence, $\bm{A} \in \mathbb{R}^{M \times M}$ is at most of rank $N-1$, 
which calls for a lower-dimensional approximation of $\bm{A}$, for instance, by \red{restricting} 
$\bm{A}$ \red{to act} on a subspace 
spanned by, say, $r$ POD modes obtained by calculating the compact SVD of $\bm{X}$,
\beq
\label{eq:svd}
\bm{X} =\bm{U}_{\bm{X}} \bm{\Sigma}_{\bm{X}} \bm{W}_{\bm{X}}^{T},
\eeq
where the superscript $T$ denotes the transpose. The truncation number $r$ is bounded from above by the rank 
of the data matrix $\bm{X}$, which is at most $N-1$.
The columns of $\bm{U}_{\bm{X}} \in \mathbb{R}^{M \times r}$  and 
the rows of $\bm{W}_{\bm{X}} \in \mathbb{R}^{\red{(N-1) \times r}}$ are orthogonal, 
and $\bm{\Sigma}_{\bm{X}} \in \mathbb{R}^{r \times r}$ is a diagonal matrix containing the nonzero singular values of $\bm{X}$.
The matrix $\bm{U}_{\bm{X}}$ contains the spatial structures of the data sequence, that is, the POD modes are given by the columns of $\bm{U}_{\bm{X}}$.
\red{Restricting $\bm{A}$ to act on the subspace spanned by $r$ POD modes gives rise to the definition of an auxiliary matrix}
\beq
\label{eq:projection}
\bm{S} := \bm{U}_{\bm{X}}^{T} \bm{A} \bm{U}_{\bm{X}} = \bm{U}_{\bm{X}}^{T} \bm{Y} \bm{W}_{\bm{X}} \bm{\Sigma}_{\bm{X}}^{-1} \in \mathbb{R}^{r \times r} \ .
\eeq
This equation is to be interpreted in a least-squares optimal sense (hence the absence of the residual), it is obtained by 
calculating $\bm{A}$ through the pseudo-inverse of $\bm{X}$, which is calculated via SVD, and subsequently 
projecting $\bm{A}$ onto the $r$-dimensional subspace spanned by the POD modes, \red{i.e. using the orthogonality of $\bm{U}_{\bm{X}}$}.  
\red{The} eigenvalues of $\bm{S}$ correspond to \red{a subset of} the non-zero eigenvalues of $\bm{A}$.
For practical purposes we summarize the SVD-based DMD algorithm \cite{Schmid2010} as follows:
\begin{itemize}

\item  Collect $N$ temporally equidistant samples $\left\{\bm{x}_{1}, \bm{x}_{2}, \bm{x}_{3}, \ldots, \bm{x}_{N}\right\}, \bm{x}_{j}\in\mathbb{R}^{M}, j \in \{1,\ldots,N \}$.

\item  Build a matrix $\bm{X} \in \mathbb{R}^{M \times(N-1)}$ out of the first $(N-1)$ snapshots, according to Eq.~\eqref{eq:X1}.

\item  Calculate the compact SVD of $\bm{X}$ according to Eq.~\eqref{eq:svd}.

\item  
Build a matrix $\bm{Y} \in \mathbb{R}^{M \times(N-1)}$ out of the
		last $(N-1)$ snapshots, according to Eq.~\eqref{eq:X2} and combine it with the matrices $\bm{U}_{\bm{X}}$
		and $\bm{W}_{\bm{X}}$ to calculate the optimal representation $\bm{S}$ of the linear mapping $\bm{A}$ in the
		orthogonal basis given by the POD modes according to Eq.~\eqref{eq:projection}.

	\item  Calculate the eigenvectors $\bm{v}_k$ and eigenvalues $\lambda_k$ of $\bm{S}$ for $k \in \{1, \hdots , r\}$.

\item Calculate the (projected) dynamic modes $\psi_{k}$
\beq
\psi_{k}=\bm{U} \bm{v}_{k} \ .
\eeq
\end{itemize}

The data vector $\bm{x}$, or, in the present context the velocity field, at time \blue{$t_s = s\Delta t$, where $s$ is an integer,} can then be approximated 
using $N' \leqslant r$ dynamic modes and their corresponding DMD eigenvalues  
\blue{
\beq
\label{eq:approx}
\bm{x}(t_s) = \bm{x}(s \Delta t) \approx \sum_{k=1}^{N'} b_{k} \lambda_k^s \psi_{k} \ ,
\eeq
}
where $b_k$ are the components of the least-squares solution of this equation at $s = 0$. \blue{The real and imaginary parts of the logarithm of the eigenvalues $\lambda_k$}
\beq
\label{eq:eigenvalues}
\omega_{k}=\frac{Im\left(\ln\left(\lambda_{k}\right)\right)}{\Delta t}, \quad \sigma_{k}=\frac{Re\left(\ln\left(\lambda_{k}\right)\right)}{\Delta t} \ 
\eeq
are the frequency and temporal growth or decay rate of the $k^{\rm th}$ dynamic mode for $k \in \{1, \hdots , r\}$, respectively. 
The accuracy of the approximation does not only depend on the number of dynamic modes used to reconstruct the 
data, it also depends on the truncation number $r$, which determines the accuracy with which the projected dynamic modes have been calculated. 
Several truncation criteria have been developed to determine a suitable value for $r$,  such as Optimal Singular Value Hard Threshold \cite{Gavish2014}, 
or to choose a number of nonzero mode coefficients from a larger number \blue{of} modes. The sparsity-promoting algorithm \cite{Jovanovic2014}, does the latter. \red{Kou and Zhang \cite{Kou2017} also developed an improved rank selection criterion for the most dominant DMD modes, which are determined based on the temporal history of each DMD mode and has better modal convergence compared to the classical DMD.} 
So far, the dynamic modes are ordered by amplitude, which may or may not result in a good reconstruction of the data, 
as the most energetic modes may not be the most dynamically relevant ones, and a mode selection criterion is required. 
Concerning low-dimensional data representation, 
sparsity-promotion \cite{Jovanovic2014} solves this impasse by minimising the least-squares error between the original and the reconstructed data over the available
set of dynamic modes, and it includes an $L_1$-penalisation term to restrict the number of active modes used for
reconstruction. As the focus here is on flow features on large spatio-temporal scales that represent statistically stationary dynamics, 
we only retain modes with eigenvalues that lie on or very close to the unit circle. The remaining modes are then ranked by frequency in ascending order, 
and we restrict our attention on the first few low-frequency modes. 
We point out that this criterion may not \blue{be} adequate for flows that feature multiple important dynamical features with
vastly different timescales, in which case either temporal filtering or a different mode selection crterion may be required. 
We will come back to this point in Sec.~\ref{sec:hc} on horizontal convection.

\subsection{Streaming DMD}
\label{sec:sdmd}

Classical DMD requires access to the entire data sequence at once, which
precludes the analysis of large datasets due to memory constraints.  This
applies to data of either a high degree of spatial or temporal complexity, 
where the former results in high spatial dimensionality (large $M$)  and
the latter requires long time series (large $N$) to capture the temporal
features of the dynamics.  Streaming DMD is a method for the calculation of the
POD-projected linear operator $\bm{S}$ based on incremental data updates that
addresses this challenge by only requiring two data samples to be held in
memory at a given time \cite{Hemati2014}. In what follows we summarize this
procedure; further details including processing steps that reduce the effects
of data contamination by noise can be found in the original work by \cite{Hemati2014}. Streaming DMD consists of two conceptual parts, a low-storage calculation of $\bm{S}$, and a scheme to update $\bm{S}$  
using new data samples based on the iterative Gram--Schmidt orthogonalization. 

Let us re-consider the data matrices $\bm{X}$ and $\bm{Y}$ defined in 
Eqs.~\eqref{eq:X1} and \eqref{eq:X2} and write Eq.~\eqref{eq:projection} as 
\beq
\label{eq:sdmd}
\bm{S} = \bm{U}_{\bm{X}}^T \bm{Y} (\bm{U}_{\bm{X}}^T \bm{X})^+ 
       = \bm{U}_{\bm{X}}^T \bm{U}_{\bm{Y}} \tilde{\bm{Y}} \tilde{\bm{X}}^+ 
       = \bm{U}_{\bm{X}}^T \bm{U}_{\bm{Y}} \tilde{\bm{Y}} \tilde{\bm{X}}^T(\tilde{\bm{X}}\tilde{\bm{X}}^T)^+ 
       = \bm{U}_{\bm{X}}^T \bm{U}_{\bm{Y}} \bm{H} \bm{G}_{\bm{X}}^+ \ ,
\eeq
where $\tilde{\bm{Y}} := \bm{U}_{\bm{Y}}^T \bm{Y} \in \mathbb{R}^{r_{\bm{Y}} \times N-1}$ and 
$\tilde{\bm{X}} := \bm{U}_{\bm{X}}^T \bm{X} \in  \mathbb{R}^{r_{\bm{X}} \times N-1}$, 
\red{are the projected data matrices, and $\bm{H} := \tilde{\bm{Y}} \tilde{\bm{X}}^T \in \mathbb{R}^{r_{\bm{Y}} \times r_{\bm{X}}}$ and 
$ \bm{G}_{\bm{X}} = \tilde{\bm{X}}\tilde{\bm{X}}^T \in \mathbb{R}^{r_{\bm{X}} \times r_{\bm{X}}}$. 
The} identity $\tilde{\bm{X}}^+ = \tilde{\bm{X}}^T(\tilde{\bm{X}} \tilde{\bm{X}}^T)^+$, 
which can be readily verified via SVD, was used in the penultimate step. That is, now both data 
matrices $\bm{X}$ and $\bm{Y}$ are projected onto orthogonal bases consisting of their 
respective left singular vectors, the POD-modes, with truncation numbers $r_{\bm{X}} \leqslant \text{rank } \bm{X}$ and $r_{\bm{Y}} \leqslant \text{rank } \bm{Y}$. 
The rearrangement carried out in the penultimate step has the advantage that 
\red{$\bm{H} \in \mathbb{R}^{r_{\bm{Y}} \times r_{\bm{X}}}$ and 
$ \bm{G}_{\bm{X}} \in \mathbb{R}^{r_{\bm{X}} \times r_{\bm{X}}}$}, 
which in itself is an improvement of classical DMD in terms of memory usage as long as $r_{\bm{X}} < M$ and $r_{\bm{Y}} < M$. 
Especially in fluid dynamics, this is often the case unless the data is very noisy. We will come back to this issue in due course.   
However, the main advantage of the formulation in Eq.~\eqref{eq:sdmd} lies in the 
fact that all matrices on the right-hand side of Eq.~\eqref{eq:sdmd} 
can be obtained incrementally from a data stream using only two samples at a time. 
The matrices $\bm{U}_{\bm{X}}$ and $\bm{U}_{\bm{Y}}$ can be calculated incrementally from 
the data stream by iterative Gram-Schmidt orthogonalization. After each 
orthogonalization step the updated orthogonal matrices are then used to 
project the sample vectors onto the respective bases, and the matrices $\bm{H}$ and $ \bm{G}_{\bm{X}}$ are
subsequently constructed from the projected sample vectors. 
More precisely, consider for instance the $k^{\rm th}$ pair of sample vectors 
$\bm{x}_k$ and $\bm{y}_k = \bm{x}_{k+1}$. The matrices $\bm{U}_{\bm{X}}$ and $\bm{U}_{\bm{Y}}$, which  have 
been constructed incrementally from the previous data samples, are now updated using 
$\bm{x}_k$ and the newly available $\bm{y}_k$. Then, $\tilde{\bm{x}}_k = \bm{U}_{\bm{X}}^T \bm{x}_k$ and
$\tilde{\bm{y}}_k = \bm{U}_{\bm{Y}}^T \bm{y}_k$ are calculated and we can update the remaining 
matrices according to 
\beq
\bm{H} = \sum_{l = 1}^k \tilde{\bm{y}}_l \tilde{\bm{x}}_l^T \qquad \text{and} \qquad \bm{G}_{\bm{X}} = \sum_{l = 1}^k \tilde{\bm{x}}_l \tilde{\bm{x}}_l^T \ . 
\eeq

Before proceeding to the calculations, a few comments are in order. First, the incremental nature of the method 
precludes the application of numerically more stable orthogonalization methods such as Householder reflections, 
and this may affect the convergence properties of the method. Second, experimental noise may result in 
a drastic decrease in computational efficiency as noise usually results in the data matrices being of high rank. 
In practice, this can be mitigated through an intermediate processing step,
as explained in detail in \cite{Hemati2014}. 
Third, we note that $\bm{G}_{\bm{X}} = \bm{\Sigma}_{\bm{X}} \bm{\Sigma}_{\bm{X}}$ 
contains the squares of the nonzero singular values of $\bm{X}$, as can be verified via SVD.

\subsection{Validation}
\label{sec:validation}
\red{
Before applying sDMD to the three aforementioned datasets, we first 
compare classical and streaming DMD implementations in terms of their respective 
memory consumption for a publicly available dataset \cite{kutz2016dynamic}
that has been extensively used for testing and validation purposes in the literature  
\cite{erichson2019rdmd,Anantharamu2019}. 
Subsequently, we use this dataset to test DMD in conjunction with a 
coarsening interpolation scheme designed to reduce the computational effort 
when analysing data of high spatial dimension $M$, as will be 
the case for turbulent flows.
}
%

Since the focus of the present work lies in the identification of the \red{large-scale features which happen to be also the} 
\red{energetically} dominant structures of the system, usually represented by one or two of the \red{dynamic modes with the highest amplitudes,} 
we do not apply any specific algorithm to \red{order the dynamic modes or to} 
determine the truncation number $r$. Instead, different values of $r$ were tested to ensure
convergence with respect to changes in the mode order and values of the lowest frequencies, resulting in $r = 30$ as a sufficient truncation number. 
\red{For data where dynamical relevance and energy content of the determined dynamic modes 
result in different mode ordering, 
more sophisticated methods such as sparsity promotion \cite{Jovanovic2014} are required to obtain good 
low-dimensional data representations.}

\subsubsection{Comparsion between sDMD and DMD}

The dataset provided in Ref.~\cite{kutz2016dynamic} consists of a time series of two-dimensional vorticity fields obtained by computer simulation of 
the wake flow behind a cylinder for Reynolds number $Re=UD/\nu = 100$, where $U$, $D$ and $\nu$
denote the free-stream velocity, the diameter of the cylinder and the kinematic
viscosity of the fluid. The dominant dynamics is governed by periodic vortex shedding, therefore it is very well suited for DMD validation. 
\red{The vortex-shedding frequency can be expressed in non-dimensional form through the Strouhal number $St = fD/U$. Here, the Strouhal number is around $St = 0.16$. }
For details on the numerical method used to generate the data we refer to the original reference \cite{kutz2016dynamic}.
In total, 150 vorticity-field samples, separated by a time interval $\Delta
t=0.2$, were analysed. 

\begin{figure}
 \centering

  \includegraphics[width=0.9\columnwidth]{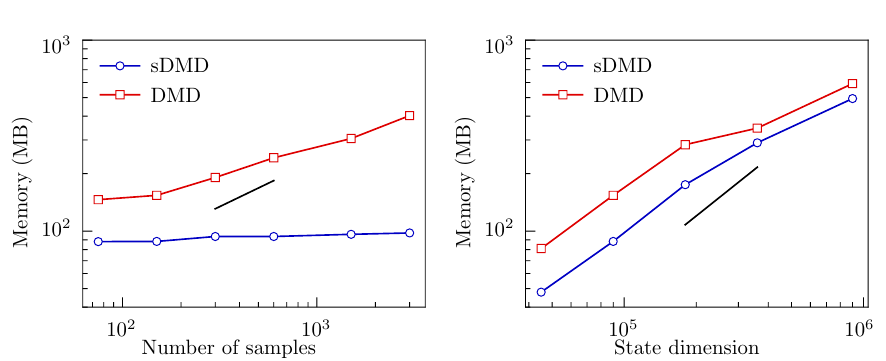}
 \put(-460, 160){\large$(a)$}
 \put(-230, 160){\large$(b)$}

 \caption{
	 Memory consumption of classical SVD-based DMD \red{(red)} and  
         streaming DMD \red{(blue)} as a function of \red{(a) the number of samples with fixed state dimension and (b) the state dimension with fixed number of samples 
	 for the vortex shedding data time series provided in Ref.~\cite{kutz2016dynamic}.
	 The solid black lines in (a) and (b) correspond to scaling exponents 0.5 and 1, respectively.} 
	} 
 \label{fig2}
\end{figure}

The memory consumption of both methods, 
\red{that is, the RAM usage of the code at a single iteration in case of sDMD 
and for the full dataset for DMD, has been assessed by two comparisons.
First we increased the number of samples for a fixed state dimension, and secondly 
increased the state dimension for a fixed number of samples.} 
In order to ensure
consistency, all tests were carried out on the same computer with an Intel
i5-8250U CPU at 1.60GHz and 8GB RAM. 
\red{We expect
the memory usage to increase with the number of data samples for DMD, 
as each additional data sample requires the same additional amount of memory. However, the 
scaling should be nonlinear, as the efficiency of the SVD is not linearly 
related to the data matrix size when the state dimension is much larger than the number of samples and as 
the memory consumption due to matrix multiplications depends on the number of data samples.  
For sDMD, the memory consumption should remain constant, as only two data samples are held in memory at a 
given time. Concerning the memory consumption as a function of the state dimension, we 
expect a linear relation for both methods.
The predictions are confirmed by the data to a good approximation, 
as can be seen in Fig.~\ref{fig2}(a) for memory consumption as a function of the 
number of data samples, and in Fig.~\ref{fig2}(b) as a function of the state dimension.
In the former case, we indeed observe nonlinear scaling of memory consumption as a function of the number of 
data samples for DMD, while the memory consumption remains constant for sDMD.
In the latter case, the memory consumption scales linearly with state dimension. Compared to DMD, it is lower by a nearly constant offset for sDMD, 
which results from the larger number of snapshots required by the DMD algorithm. For DMD all 150 snapshots are stored in memory, 
while sDMD requires only two.
Information on computational time and memory usage for DMD and sDMD calculations for 150 snapshots on 90000 grid points is provided in table \ref{table1}. 
}

\subsubsection{Coarse interpolation for the analysis of high-dimensional data}
\begin{figure}
 \centering
 \includegraphics[width=1\columnwidth]{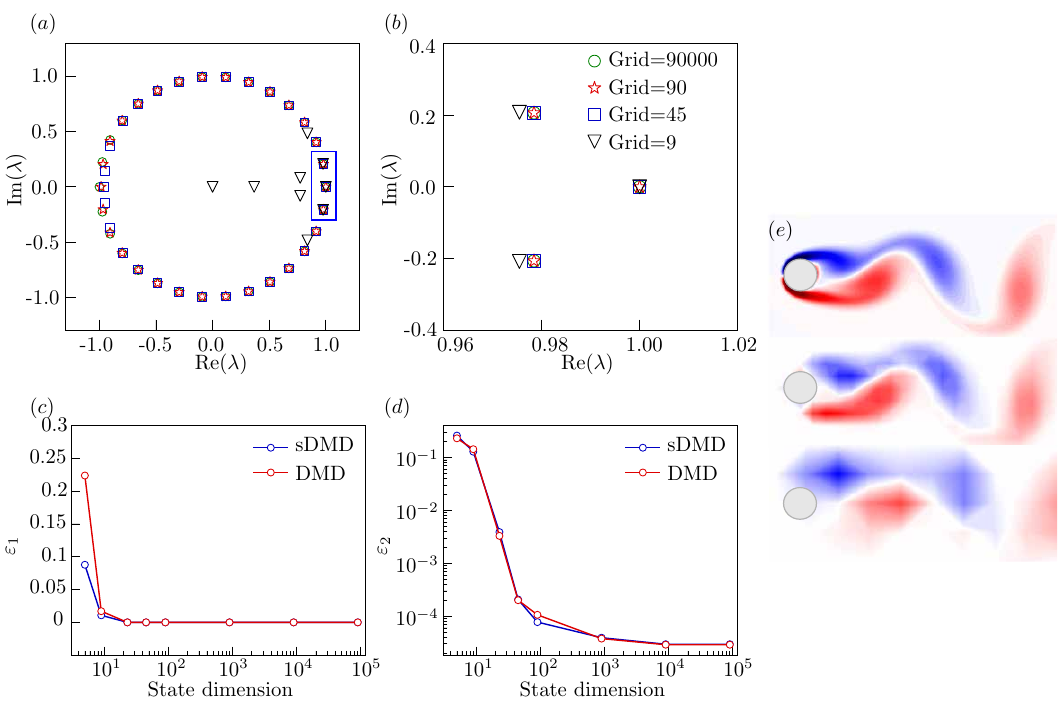}
	\caption{ 
   	(a) Streaming DMD eigenvalues for the original data (green circles) and after different degrees of downsampling. 
	(b) Magnification of the blue region in (a).
	(c) Error of the Strouhal number as a function of the \red{state dimension for the downsampled data} 
	for classical DMD and sDMD.
	(d) The time-averaged reconstruction error as defined in Eq.~\eqref{eq:recerr} as a function of the \red{state dimension for the downsampled data} 
	for classical DMD and sDMD. 
        (e) The instantaneous flow field on 90000, 900 and 100 grid points, respectively, from top to bottom. 
	 }
 \label{fig3}
\end{figure}

Numerical simulations of highly turbulent flows require fine computational grids 
to accurately resolve the dynamics at the small scales. This 
is not necessarily always due to a need to precisely measure small-scale quantities such as 
dissipation \red{or correlation functions of high order}, it is also a requirement for 
numerical stability.  
The required large number of grid points results in a high memory load even for a single 
sample, which quickly becomes prohibitive even for sDMD. 
This calls for a reliable downsampling strategy to interpolate the data
on coarser grids, \red{in particular when the focus is on large-scale structures.}
In what follows we analyse the robustness of sDMD with respect to different degrees
of spatial downsampling, using the same vortex shedding dataset of the previous subsection. 
The downsampling was carried out by successively decreasing the original number of grid points 
\red{uniformly, that is by merging nearby grid points}, beginning
with 90000 grid points down to a minimum of 5 grid points. The effect of downsampling is assessed by considering 
two observables, the DMD eigenvalues and the time-averaged reconstruction error, defined as 
\beq
\label{eq:recerr}
\blue{
\varepsilon_2 := \left \langle \|\bm{v}(t_s) - \sum_{k=1}^{N'} b_{k} \lambda_k^s \psi_{k} \|_2 \right \rangle \ ,
}
\eeq
where $\bm{v}$ is the \red{downsampled} vorticity field here, and the angled brackets denote
a time average.  
The results are summarized in Fig.~\ref{fig3}, with
Fig.~\ref{fig3}(a) and \ref{fig3}(b) showing the streaming DMD eigenvalues for the
original data and after different degrees of downsampling, Fig.~\ref{fig3}(c)
the reconstruction error as as a function of the \red{state dimension} 
for classical DMD and sDMD, and Fig.~\ref{fig3}(d) presenting visualisations of a sample of
the reconstructed vorticity fields after different degrees of downsampling.  A
number of observations can be made from Fig.~\ref{fig3}.  The DMD eigenvalues,
\red{which need to lie on the unit circle as the dynamics are nonlinear
and statistically stationary \cite{horn2017}, }
are remarkably robust under the downsampling procedure.
\blue{This can be expected as the Koopman operator
can be approximated for any observable function. Thus, a coarsening of the grid represents a change in the observable, which should not have a large impact on the eigenfunctions.}
As can be seen from
the data shown in Fig.~\ref{fig3}(a) and \ref{fig3}(b), a reduction by three orders of
magnitude in the \red{state dimension} 
results in almost the same values for the DMD eigenvalues. Significant qualitative differences in the eigenvalues occur only
after drastic downsampling from 90000 to less than 10 data points. A more
quantitative comparison is achieved by considering the difference
$\varepsilon_1$ between the Strouhal number and the dimensionless frequency of
the second dynamic mode as a function of the \red{state dimension}
presented in Fig.~\ref{fig3}(c). As can be seen from the figure, the Strouhal number is
reproduced very accurately using only 25 data points.  This is particularly
striking in view of the unsurprisingly large reconstruction error $\varepsilon_2$
of order $10^{-3}$ to $10^{-2}$, for the corresponding downsampled data, as
shown in Fig.~\ref{fig3}(d).  According to the data presented in the figure,
converged results for the reconstruction of the full vorticity field requires
a \red{state dimension}
of least 9000 points. The finite residual for higher resolved data is then due to truncation in the DMD algorithm. 
The visualisations of the reconstructed vorticity fields in Fig.~\ref{fig3}(e) give a visual impression of the effect the downsampling has on the
reconstructed data. As expected, the large-scale spatial coherence is still
present in the downsampled data.  Since the focus is on the detection of
large-scale coherent structures, like the vortex street in this case, and since
the coarsening interpolation results in the removal of small-scale spatial
structures, the downsampling  has very little effect on the results, as
expected.

\section{Results}
\label{sec:results}
Having validated our implementation of sDMD \red{in conjunction with downsampling} 
on publicly available data, 
we now apply the method to three different flows, rapidly rotating Rayleigh--B\'enard
convection (RBC), horizontal convection (HC), and the asymptotic suction
boundary layer (ASBL). We chose these three examples in order to demonstrate
sDMD to be a useful tool for the analysis of different turbulent flows in terms
of their main spatio-temporal structure.  

In rapidly rotating RBC, the anticyclonic circulation in the bulk is surrounded
by a cyclonic layer close to the horizontal cell walls, and the aim is to
identify this large-scale flow pattern. Horizontal convection lends itself well
as a test case for the distinction of different spatio-temporal structures, as
the dynamics is largely governed by two instabilities that operate on different
time scales. The Rayleigh--Taylor instability leads to fast periodic plume
generation close to the boundary while an oscillatory instability in the bulk
results in much slower periodic dynamics in the bulk. The respective
frequencies associated with these two processes differ by an order of
magnitude.  Similar to canonical wall-bounded parallel shear flows and
spatially developing BLs, the  ASBL features long-lived large-scale coherent
motion. Here the aim is to identify the corresponding spatio-temporal
structure. The slow dynamics requires very long time series, which makes this
example particularly suitable for the application of sDMD.  

All datasets were obtained by direct numerical simulation at parameter values
corresponding to turbulent flow. Further details on the numerical methods and
parameter values will be given in the following subsections. 
\red{A summary of computational details such as wall time and memory consumption for the 
DMD or sDMD calculations is provided in table \ref{table1} for all datasets.}

\begin{table}
    \centering
    \begin{tabular}{p{2.5cm}p{1.3cm}p{2cm}p{2cm}p{4.3cm}p{1cm}}
        \hline
	    dataset           & wall time [sec] & memory DMD [MB] & memory sDMD [MB] & CPU                        & RAM [GB] \\
        \hline
	    RRB-$\Ra=10^8$    & 940            &   -               & 6201             & Intel i5-4440 3.10GHz  & 16  \\
	    RRB-$\Ra=10^9$    & 2167           &   -               & 4230             & Intel i5-9400 2.90GHz  & 16  \\
	    HC-slow           & 453            &   -               & 2060             & Intel i5-9400 2.90GHz  & 16  \\
	    HC-fast           & 2              &   -               &  37              & Intel i5-9400 2.90GHz  & 16  \\
	    ASBL              & 44             &   -               &  204             & Intel i7-8550U 1.80GHz & 16  \\
	    cylinder wake     & 8              & 154               &  88              & Intel i5-8250U 1.60GHz & 8  \\

        \hline       
    \end{tabular}
	\caption{
		\red{
		Memory requirements for all datasets. For the cylinder flow dataset discussed in sec.~\ref{sec:validation}, 
		150 data snapshots on 90000 grid points have been used, and the wall time refers to the sDMD calculation. 
		}
        }
    \label{table1}
\end{table}

\subsection{Rapidly Rotating Rayleigh--B\'enard Convection}
\label{sec:rbc}
\subsubsection{Fluid structures}
In rotating Rayleigh--B\'enard convection, a fluid is confined between a heated
bottom plate and a cooled top plate and is rotated around a vertical axis. It
is a paradigmatic problem to study many geophysical and astrophysical phenomena
in the laboratory, e.g. convective motion occurring in the oceans, the
atmosphere, in the outer layer of stars, or in the metallic core of planets.
In rotating RBC laboratory experiment, the fluid is laterally confined.  The
centrifugal force can be neglected, provided the Froude number is small, and
then only the Coriolis force is considered.  The interplay of the occurring
buoyancy and Coriolis forces, however, may yield highly complex flows with very
distinct flow structures whose nature strongly depends on the control
parameters.  Without rotation or with slow rotation, a distinct feature of
turbulent RBC is the emergence of the Large-Scale Circulation (LSC) of fluid.
For rapid rotation, however, a mean flow with cyclonic azimuthal velocity near the boundary,
the Boundary Zonal Flow (BZF), develops close to the side walls, surrounding a
core region of anticyclonic mean flow.  The viscous Ekman BLs near the plates
induce an anticyclonic circulation with radial outflow in horizontal planes,
which is balanced by the vertical velocity in a thin annular region near the
sidewall, where cyclonic vorticity is concentrated.  The Taylor--Proudman
effect induced by rapid rotation tends to homogenize the flow in the vertical
direction, resulting in an anticyclonic mean flow in the core region throughout the height. The temperature pattern near the vertical wall, however, moves
anticyclonically within the BZF and is likely connected to the thin anticyclonic Ekman
layers at the top and bottom plates. The interesting part of the BZF flow structure is, that it has an organized and predominant pattern as mean flow, although the instantaneous flow is turbulent in the whole domain, and consists of active and complex vortex motion. This grants sDMD big potential, as it can find the dominant modes quickly and thus reconstruct the global statistics at very low cost.
In addition, the BZF has special drift feature, which could test how well sDMD could capture the temporal evolution of a flow. Thus the aim here is to recover the BZF via sDMD.

\subsubsection{Dynamic equations \& control parameters}
We  consider  RBC  in  a  vertical  cylinder  rotating  with  uniform angular
velocity $\Omega$ about  the  vertical  axis. The  governing  equations  of  the
problem  are  the  incompressible  Navier--Stokes  equations  in  the
Oberbeck--Boussinesq approximation,  coupled  with  the  temperature  equation, given here in dimensionless form
\begin{align}
	\label{eq:ns}
	\partial_t \bm{u} + (\bm{u}\cdot\nabla)\bm{u} + \nabla p & = \sqrt{\Pran/\Ra}\nabla^2\bm{u} -\Ro^{-1} \hat{z} \times \bm{u} + T\hat{z}, \\
	\label{eq:energy}
	\partial_t T+ (\bm{u}\cdot\nabla) T &= \sqrt{1/(\Pran\Ra)}\nabla^2 T,\\
	\label{eq:incomp}
	\nabla\cdot\bm{u}&=0.
\end{align}
%
\red{which are nondimensionalized by using the fluid layer height $H$, the temperature differences between heated bottom and cooled top plates ${\Delta=T_+-T_-}$, and the free-fall velocity
$u_{ff}=\sqrt{\alpha g\Delta H}$, with $\alpha$ denoting the isobaric expansion coefficient, $g$ the acceleration due to gravity}. 
The Rayleigh number, $\Ra$, describes the strength of the thermal buoyancy force,
the Prandtl number, $\Pr$, the ratio of viscosity and diffusivity, and the
convective Rossby number, $\Ro$, is a measure for the rotation rate. They are
defined as
\beq
\Ra \equiv \alpha g \Delta H^3/\red{(}{\kappa\nu}\red{)},\qquad \Pran \equiv \nu/\kappa, \qquad \Ro \equiv \sqrt{\alpha g\Delta H}/\red{(}{2\Omega H}\red{)},
\eeq
where $\kappa$ the thermal diffusivity, $\nu$ the
kinematic viscosity, and $\Omega$ the angular rotation speed.  Equations
\eqref{eq:ns}-\eqref{eq:incomp} were stepped forward in time using the
finite-volume code {\sc goldfish} \cite{shishkina2015,Kooij2018,Zhang2020}.
For  the  temperature  we  impose Dirichlet  boundary  conditions (isothermal) on  the top and bottom  plates and Neumann conditions (adiabatic)  on  the lateral  walls.  
All  boundaries are  assumed  to  be  impenetrable  and no-slip,  i.~e.~the  velocity field vanishes at all boundaries.

\begin{figure}
	\begin{center}
        \includegraphics[width=0.7\columnwidth]{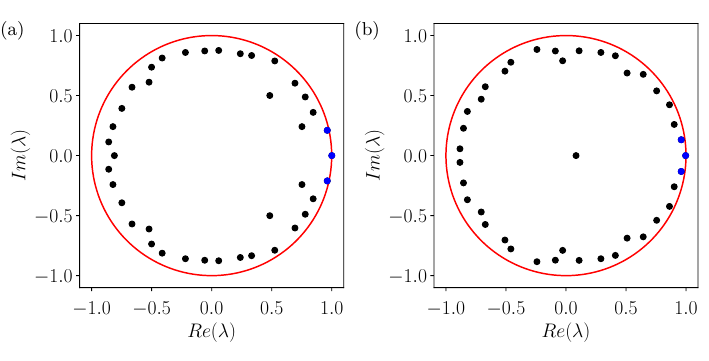} 
	\end{center}
        \caption{DMD eigenvalues for rapidly rotating Rayleigh-B\'enard convection. (a) $\Ra = 10^8$, (b) $\Ra = 10^9$. The eigenvalues shown in blue (three) correspond to the 
	modes used in the reconstruction shown in Fig.~\ref{fig4}.}
	\label{fig:RRBC-spectra}
\end{figure}

\begin{figure*}
 \unitlength1truecm
 \begin{picture}(17, 6)
 \put(0,0){\includegraphics[width=0.95\columnwidth]{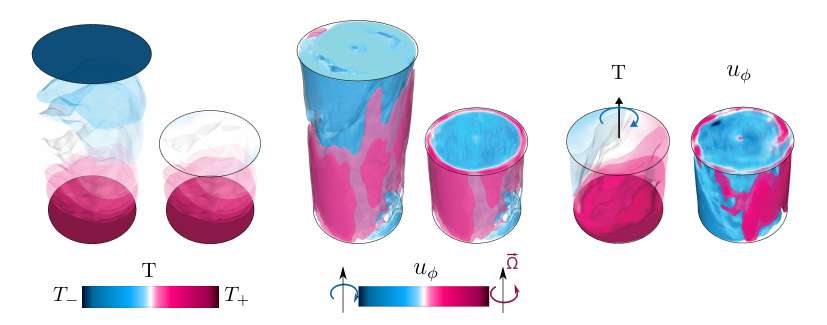}}
 \put(0, 5.5){$(a)$}
 \put(5.5, 5.5){$(b)$}
 \put(11, 5.5){$(c)$}
 \end{picture}
	\caption{The first two dynamic modes for $\Pran=0.8$, $\Ra = 10^8$ and $\Ro=0.1$. 
	$(a)$ The first mode of the anticyclonic drifting temperature field in the full cell (left) and in the bottom half of the cell (right). 
 $(b)$ The first mode of the azimuthal velocity field  in the full cell (left) and in the bottom half of the cell (right). 
	$(c)$ The second mode of the temperature (left) and the azimuthal velocity field (right) in the bottom half of the cell.}
 \label{fig4}
\end{figure*}

\subsubsection{Numerical details}
We consider datasets for two different Rayleigh numbers, $\Ra=10^8$ and
$\Ra=10^9$, the remaining control parameters are $\Pran=0.8, \Ro=0.1$.  The
resolution of the original datasets is $N_{r} \times N_{\phi} \times N_{z}=100
\times 256 \times 380$ for $\Ra = 10^8$ and $N_{r} \times N_{\phi} \times
N_{z}=192 \times 512 \times 820$ for $\Ra = 10^9$, according to
\cite{Shishkina2010}, where $N_r$, $N_\phi$ and $N_z$ denote the number of grid
points in radial, azimuthal and vertical direction, respectively. Grid nodes are non-equidistant in both the radial and vertical directions, being clustered near the boundaries to resolve thermal and velocity boundary layers \cite{Zhang2021}.  The velocity
fields from both datasets are sampled for a time period of 200 free-fall time
units with a sampling interval of of $\Delta t = 1.25$,  resulting in 160
samples in total.  Both datasets are spatially downsampled for the sDMD
analysis, by a four-fold and a 16-fold reduction in the number of data points,
respectively, resulting in a spatial resolution of $N_{r} \times N_{\phi}
\times N_{z}=100 \times 128 \times 190$ for $\Ra = 10^8$ and $N_{r} \times
N_{\phi} \times N_{z}=96 \times 128 \times 410$ for $\Ra = 10^9$. The
truncation number is $r=40$ in both cases.  In what follows we first
describe \red{eigenvalue spectra and} the generic spatial features that can be extracted with the first few dynamic
mode for the case $\Ra = 10^8$. Subsequently, we consider the temporal features
for both $\Ra = 10^8$ and $\Ra = 10^9$, \red{providing a quantitative comparison of the 
time scales associated with the global structure obtained directly from the DNS data and calculated from the 
lowest DMD frequencies}. 

\subsubsection{Streaming DMD}

The DMD eigenvalues obtained for the two cases $\Ra = 10^8$ and $\Ra = 10^9$ are presented 
in Fig.~\ref{fig:RRBC-spectra} (a) and (b), respectively. As can be seen from the data shown in 
the figure, the eigenvalues lie on or close to the unit circle for a truncation number $r = 40$. 
We only consider converged modes corresponding to eigenvalues on the unit circle, 
that is, the mean flow and the dynamic mode corresponding to the next lowest frequency indicated by the blue dots. 
The first two dynamic modes obtained from the $\Ra = 10^8$-dataset are
visualised in Fig.~\ref{fig4} in terms of temperature and azimuthal velocity.
The temperature field of the first dynamic mode shown in Fig.~\ref{fig4}(a)
resembles the mean temperature profile, the corresponding azimuthal velocity
field (Fig.~\ref{fig4}b) consists of anticyclonic motion in the bulk and
cyclonic motion close to the sidewall.  As expected, the first dominant mode
corresponds to a base or mean flow.  However, this mode is evidently
dynamically not important without temporal change.  The principal mode is the
second one, presented in Fig.~\ref{fig4}(c), and the BZF \cite{Zhang2020} is
clearly visible, both in the temperature (left) and azimuthal velocity (right).  

\begin{figure}
 \unitlength1truecm
 \begin{picture}(15, 6.7)
 \put(2,0){
 \put(0.5, 0){\includegraphics[width=12cm]{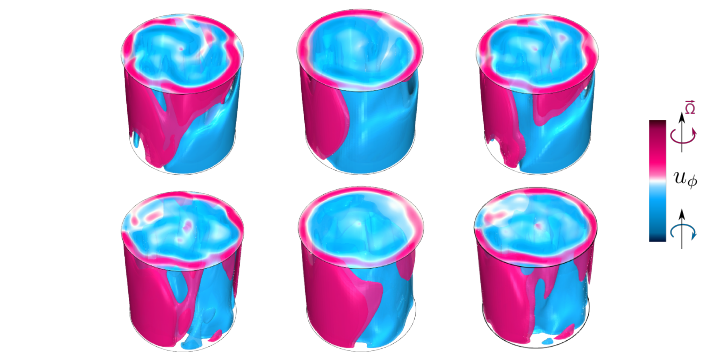}}
 \put(2, 5.8){$(a)$}
 \put(5.1, 5.8){$(b)$}
 \put(8.2, 5.8){$(c)$}
 \put(2, 2.7){$(d)$}
 \put(5.1, 2.7){$(e)$}
 \put(8.2, 2.7){$(f)$}
 \put(0.3, 4.8){\text{snapshot 1}}
 \put(0.3, 1.5){\text{snapshot 2}}
 \put(2.7, 6.2){\text{Original}}
 \put(5.8, 6.2){\text{2 Modes}}
 \put(8.9, 6.2){\text{5 Modes}}}
\end{picture}
 \caption{
	 Reconstructed azimuthal velocity field for two velocity-field samples for $\Pran=0.8$, $\Ra = 10^8$ and $\Ro=0.1$.
 $(a),(d)$ original field,
 $(b),(e)$ reconstruction with two dynamic modes,
 $(c),(f)$ reconstruction with five dynamic modes.
 }
 \label{fig6}
\end{figure}

Even though the flow is turbulent, its large-scale spatial structure can be
reconstructed nicely with only a few modes as demonstrated by the visualisation
of the azimuthal velocity in the lower half of the RBC-cell presented in
Fig.~\ref{fig6}. We point out that much of the small-scale dynamics and
thereby accuracy in the representation is lost through the downsampling
procedure, and applying streaming DMD without coarse interpolation but with
larger memory consumption may be advisable if the focus is on a more detailed
reconstruction of the flow.  Here, we focus only on the large-scale structure.
The comparison is carried out for two velocity fields which have been sampled
about 30 free-fall times apart in order to guarantee sufficiently decorrelated
samples.  The originals are shown in Fig.~\ref{fig6}(a) and \ref{fig6}(d),
respectively.  Figure~\ref{fig6}(b) and \ref{fig6}(e) contain the reconstructions using
the first two modes, and Fig.~\ref{fig6}(c) and \ref{fig6}(f) the reconstructions from
the first five modes for the two samples, respectively.  These examples
demonstrate consistently that even though the main features of the flow can be
captured by the mean flow and the BZF, a fair amount of detail is missing and
its inclusion requires a few more modes.  A much better reconstruction can be
achieved with as little as five modes. 

\begin{figure}[h!]
 \unitlength1truecm
 \begin{picture}(17, 5)
 \put(0,0){\includegraphics[width=0.95\columnwidth]{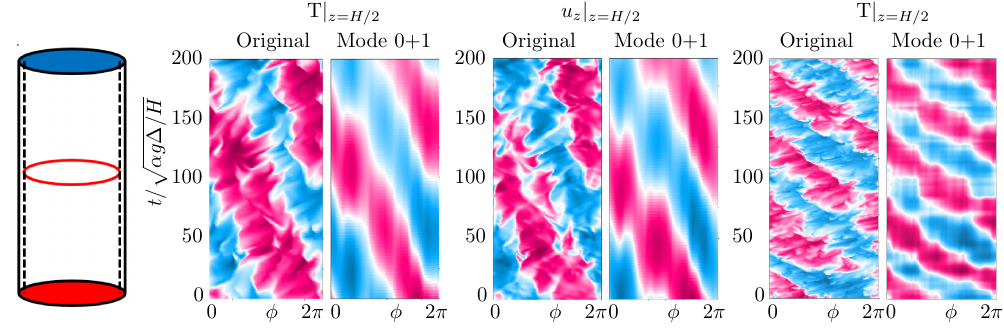}}
 \put(-0.2, 5.3){$(a)$}
 \put(2.3, 5.3){$(b)$}
 \put(7.5, 5.3){$(c)$}
 \put(12.2, 5.3){$(d)$}
 \end{picture}
\caption{
Time evolution of temperature and vertical velocity.
$(a)$ Schematic setup. The red circle indicates the location where temperature and velocity were measured.
$(b)$ temperature field and $(c)$ vertical velocity field for $\Ra = 10^8$, and
$(d)$ temperature field for $\Ra = 10^9$. The original fields are shown in the left panels and the right panels correspond to the reconstructed field using two dynamic modes. 
The color scale varies from minimum values indicated in blue to maximum values indicated in magenta for the respective fields, given by temperatures at the top and bottom plates in  $(b)$ and $(d)$, 
	and $[-u_{ff}/2, u_{ff}/2]$ with $u_{ff} \equiv \sqrt{\alpha g\Delta H}$ being the free-fall velocity in $(c)$.}
 \label{fig5}
\end{figure}

Having discussed the identification of the dominant spatial feature of the
flow, the BZF, we now focus on its temporal structure.  Figure~\ref{fig5}
presents spatio-temporally resolved diagrams of the dynamics in a ring located
at half-height $z = H/2$ and at radial location $r=r_{u_\phi^\text{max}}$,
where the maximum azimuthal velocity is observed, as indicated by the red
circle in the schematic drawing shown in Fig.~\ref{fig5}(a).    The time
evolution of the temperature and vertical velocity fields of the $\Ra =
10^8$-dataset are presented in Fig.~\ref{fig5}(b) and \ref{fig5}(c), respectively,
while Fig.~\ref{fig5}(d) corresponds to the time-evolution of the temperature
field at $\Ra = 10^9$.  The original data is shown in the left panels of the
respective visualisations and the data reconstructed from the first two dynamic
modes is shown in the right panels. Visual comparison of the left and right
panels confirms again the zonal flow pattern can be clearly captured with
only the first two dynamic modes.  Furthermore, the
visualisations clearly identify the BZF as a travelling wave with strongly
correlated temperature and vertical velocity fields as can be seen by
comparison of Fig.~\ref{fig5}(b) and ~\ref{fig5}(c).  The travelling wave structure of
the BZF is also present at higher $\Ra$, as can be seen in
Fig.~\ref{fig5}(d).  As such, it seems to be a robust feature of the BZF in
the Rayleigh-number range considered here. However, according to the
visualisation the dynamics appear to be slightly more complex at $\Ra= 10^9$
than at $\Ra= 10^8$, hence it remains to be seen to what extent the travelling
wave dynamics persist with increasing $\Ra$.

\red{We now provide a qualitative comparison of time scales between the full data and a reconstruction using only the 
mean flow and the dynamic modes representing statistically stationary dynamics on the longest time scales 
presented qualitatively in Fig.~\ref{fig5}. For the case $\Ra = 10^8$, the BZF period obtained from the data is 
74 free-fall time units, which compares well with the period of 72.96 free-fall time units 
corresponding to the second DMD mode. The relative error between the two amounts to 1.4 $\%$. The comparison at higher 
Rayleigh number for the case $\Ra = 10^9$ results in a larger relative error, the period obtained from the data 
of 53 free-fall time units compares with a relative error of 8.5 $\%$ to the DMD-period of 57.5 free-fall time units. The latter has been calculated using the expression for DMD frequencies in Eq.~\eqref{eq:eigenvalues}.} 

In summary, the most prominent spatio-temporal features of rapidly rotating RBC
can be identified through sDMD, with the BZF emerging as the dominant dynamic
mode.  The cyclonic motion of the fluid reflected in the azimuthal velocity and
the anticyclonic motion of the flow pattern reflected in the temperature and
vertical velocity as well as their frequencies are fully reproduced by only the
first two dynamic modes.  These results firmly establish sDMD as a powerful
tool for the extraction of dominant coherent structures in turbulent rapidly
rotating RBC.

\red{Large-scale structures in RBC at $\Ra = 10^7$ and $\Pr = 0.7$ without
rotation and in a cubic domain have recently been identified through Koopman
analysis \cite{giannakis2018koopman}.  In cubic geometry, eight LSC states 
occur in RBC, four long-lived diagonal configurations
with sojourn times of $O(1000)$ free-fall time units and   four wall-aligned
configurations, which are visited shortly during transitions between the
diagonal configurations.  All LSC states have been reconstructed using three Koopman
modes, a pair of complex conjugate modes whose frequency should approximate the
time scale of one cycle through all four diagonal configurations, and a real
mode representing the mean flow. In this context, the present results suggest
that
a  representation of the BZF \blue{through Koopman eigenfunctions should also } be possible.  }


\subsection{Horizontal Convection}
\label{sec:hc}
\subsubsection{Fluid structures}
\begin{figure}
 \centering
 \includegraphics[width=0.95\columnwidth]{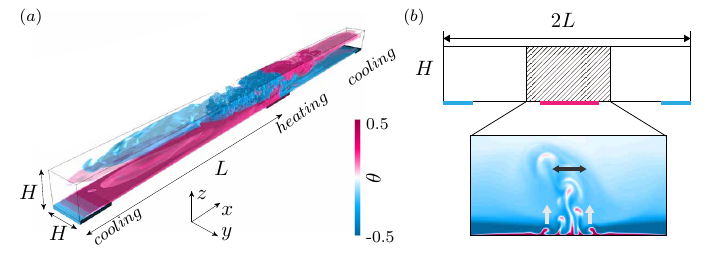}
	\caption{ 
		Sketch of (a) HC adapted from \cite{Reiter2020} and (b) front
		view of the setup. Only the shaded area (b) is used for sDMD. 
		The inset shows a snapshot of the temperature
		field for $Ra = 10^{11}$ and $Pr = 10$. There, the grey arrows
		indicate the motion of the periodically detaching plumes; the
		dark arrow indicates the oscillatory motion inside the bulk
		region.
		}
 \label{fig7}
\end{figure}

Horizontal convection (HC), similarly to RBC, is driven by thermal buoyancy.
However, in HC heating and cooling are applied to different parts of the same
horizontal surface. In our case, the heated plate is located in the center and
the cooled plates are placed at both ends, as shown in Fig.~\ref{fig7}(a).
This setup is relevant for many geophysical and astrophysical flows
\cite{Scott2001, Spiegel1971} and engineering applications \cite{Gramberg2007},
in particular concerning the large-scale overturning circulation of the ocean as heat
is supplied to and removed from the ocean predominantly through its upper
surface, where the ocean contacts the atmosphere. The dimensionless control parameters are
similar to RBC, that is the Rayleigh number, the Prandtl
number and the aspect ratio $\Gamma$,
\begin{eqnarray*}
\qquad \quad
Ra\equiv \alpha g \Delta L^3/(\kappa\nu), \qquad Pr\equiv\nu/\kappa, \qquad
\Gamma\equiv L/H=10,
\end{eqnarray*}
where the characteristic length scale $L$ is the half-cell length. 
The governing equations are again the incompressible Navier--Stokes equations in the Oberbeck--Boussinesq approximation, and a 
temperature equation stated in Eqs.~\eqref{eq:ns}-\eqref{eq:incomp}, but without the Coriolis term in the momentum equation. 

For the parameters $\Ra = 10^{11}$ and $\Pran = 10$ it was observed that
sheared plumes, originated by a Rayleigh--Taylor instability, periodically
arise above the heated plate and travel towards the center \cite{Reiter2020}.
However, another time-dependent feature that emerges is the oscillatory
instability that breaks symmetry inside the bulk region, see Fig.~\ref{fig7}(b).  So there is a fast periodic emission of thermal plumes close to the
boundary and a slow periodic oscillation in the bulk region. 
That is, the horizontal convection has coexisting dynamics on very different time scales.
Streaming DMD is used in conjunction with temporal filtering to  provide separate low-dimensional reconstructions of these coexisting dynamics.
Temporal filtering is required here 
\blue{for reasons of computational efficiency. In principle, it is possible to extract both time scales from a single 
analysis. However, the dataset would become very large and the calculations slow. This is because the 
snapshot spacing must be small enough to detect the small period and we need to process a large number of snapshots along a trajectory long enough to capture the large period. To obtain converged results in particular for the larger period, the truncation number must be increased accordingly, resulting in much larger matrices to be processed at each iteration step.  }

\subsubsection{Dynamic equations \& numerical details}
The dataset consists of velocity fields obtained in the DNS for a rectangular
geometry, as shown in the schematic drawing in Fig.~\ref{fig7}(a).  The
temperature boundary conditions at the bottom plate are $\theta =0.5$ for $0
\leq x \leq 0.1$ and $\theta =-0.5$ for $0.9 \leq x \leq 1$, all the other
walls are adiabatic.  No-slip boundary conditions are imposed at all walls for
the velocity field.  The calculations were carried out using the {\sc goldfish}
code, as in the previous section.  Further details can be found in
\cite{Reiter2020}.  The original grid is $N_{x} \times N_{y} \times N_{z}=1026
\times 66 \times 98$, where $N_x$, $N_y$, and $N_z$, denote the number of grid
points in the mean-flow $x$-direction, the spanwise $y$-direction and the
$z$-direction, which is normal to the heated and cooled bottom plates,
respectively.  Though, since plumes and oscillations are concentrated above the
heated plate, we extract only the dynamically most important data inside the shaded domain, shown in
Fig.~\ref{fig7}(b), with  $N_{x} \times N_{y} \times N_{z}=200 \times
66\times 98$. The truncation number $r$ is set to 80 to ensure the dominant
modes can be captured properly.  Since the plume emission motion is more than
ten times faster than the oscillatory flow, a small time interval is needed to
capture the fast plume emission while a large number of velocity-field samples
is required to simultaneously identify the slow oscillations. To save
computational resources, 
we decouple the two tasks and use two datasets
comprised of 200 snapshots each, sampled at different time intervals: $0.1$
free-fall time units for the fast plume emission and $0.5$ free-fall time units
for the slow oscillatory flow. \blue{As discussed in the previous section, it is in principle possible to extract both phenomena from a single dataset.}

\begin{figure}
	\begin{center}
        \includegraphics[width=0.7\columnwidth]{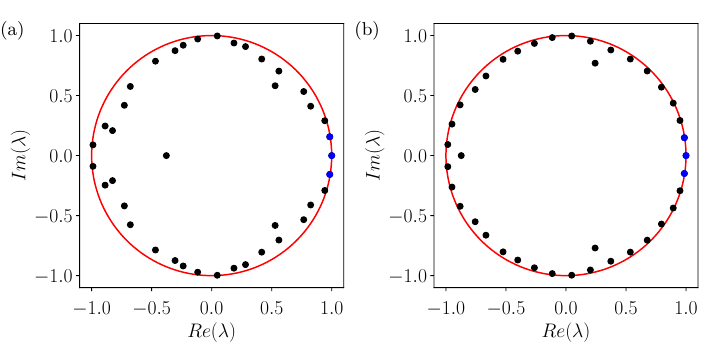} 
	\end{center}
	\caption{\red{Streaming DMD eigenvalues for horizontal convection with (a) fast plume emission and (b) slow oscillatory flow. The eigenvalues shown in blue (three) corresponds to the modes used in the reconstruction shown in Fig.~\ref{fig8}}}
	\label{fig:HC-spectra}
\end{figure}

\begin{figure}
 \centering
 \includegraphics[width=0.7\columnwidth]{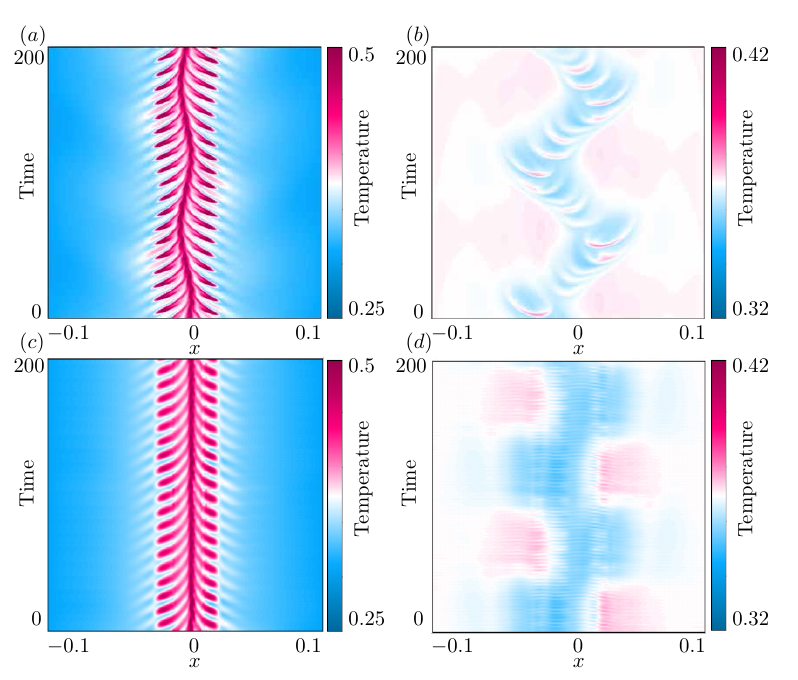}
	\caption{
	 (a,~b) Time evolution of the temperature of the original flow at a horizontal slice located at $y=H/2$ and
	 at the height (a) $z=0.1H$, to capture plume emission, and
	 (b) $z=0.8H$, to capture oscillations. (c,~d) Time evolution of the
	 temperature of the reconstructed field with the first $2$ dominant
	 modes. It is noted that
	 the reconstruction (c) and (d) are based on different snapshot
	 intervals.
	 }
 \label{fig8}
\end{figure}

\subsubsection{Streaming DMD}
\red{
Figure~\ref{fig:HC-spectra} presents the DMD spectra for fast plume emission in panel (a) and the slow oscillatory flow in panel (b). 
Similar to the RRBC DMD spectra shown in Fig.~\ref{fig:RRBC-spectra}, the eigenvalues lie on or close to the unit circle. In both cases 
the focus is on the slow dynamics in the respective datasets obtained by the two sampling procedure outlined in the previous section, 
we focus on the lowest obtained frequencies. The first five low-frequency modes lie in fact on the unit circle for both cases, 
with the data sampled at larger intervals shown in \ref{fig:HC-spectra}(b) being converged at higher frequencies as well.}

The temporal structure of the original temperature field and the temperature
field reconstructed from the first two dynamic modes is shown in
Fig.~\ref{fig8} using horizontal slices located that the spanwise middle of
the domain, $y = H/2$, and at different heights.  Figure~\ref{fig8}(a) and ~\ref{fig8}(b)
contain visualisations of the original field at $z=0.1H$, to capture fast plume
emission, and at $z=0.8H$, to capture slow oscillations, respectively, and
Fig.~\ref{fig8}(c) and ~\ref{fig8}(d) present the corresponding reconstructions.  A
visual comparison of the original and the reconstructed fields qualitatively
shows that sDMD can clearly distinguish the two dominant spatio-temporal
structures, with the first two dynamic modes identifying the fast motion of the
plume emission for the dataset sampled at $0.1$ free-fall time units
(Fig.~\ref{fig8}(a) and ~\ref{fig8}(c)), and the first two dynamic modes  capturing the
slow oscillatory mode for the dataset sampled at $0.5$ free-fall time units
(Fig.~\ref{fig8}(b) and ~\ref{fig8}(d)).
The frequencies obtained from the DNS data and the sDMD calculations are compared with the DMD frequencies 
\blue{calculated according to Eq.~\eqref{eq:eigenvalues}}.
The period of the first dynamic mode is \red{16.98} 
free-fall time units \blue{according to Eq.~\eqref{eq:eigenvalues}}, which matches \blue{very well} the period of the slow oscillation observed in the original dataset
\red{measured to be 16.8 free-fall time units}. 
The second dominant mode has a period of {1.58} 
free-fall time units \blue{according to Eq.~\eqref{eq:eigenvalues}}, which fits the period of \red{1.6 
free-fall time units of} the fast plume emission determined from the original DNS data. 
\red{In both cases, the relative error between the time scales obtained from the full data and via DMD is about 1 $\%$.}
The agreement between the sDMD results and the DNS data, and the distinct
identification and separation of the two dominant spatio-temporal structures
with frequencies that differ by an order of magnitude, gives further confidence
in the capability of DMD to capture the relevant processes, be it in the
temporal or spatial framework.

\subsection{Asymptotic Suction Boundary Layer (ASBL)}
\label{sec:asbl}
\subsubsection{Fluid flow}
\begin{figure}
 \centering
 \includegraphics[width=.5\columnwidth]{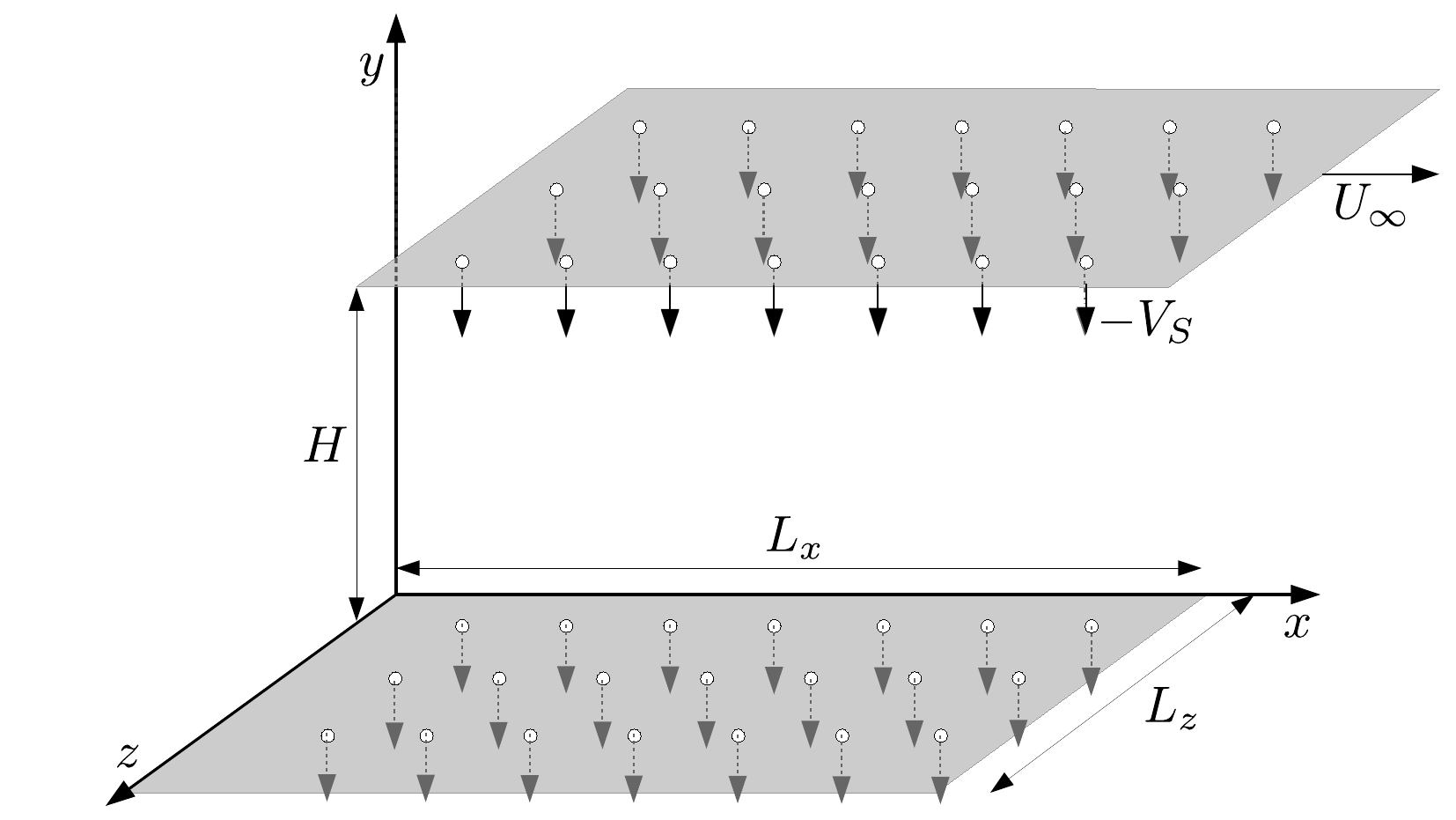}
 \caption{Schematic drawing of the asymptotic suction boundary layer in numerical simulations. 
	The lower plate is stationary and the fluid is set in motion by the 
	upper plate that moves in $x$-direction with velocity $U_\infty$, representing the free-stream velocity of 
	the emulated open flow. Fluid is removed through a porous bottom plate with velocity $V_S$, to guarantee conservation 
	of mass, fluid enters the system at the same speed through a porous top plate. In numerical simulations, this is 
	realised uniformly through boundary conditions on the wall-normal component of the velocity field. 
 }
 \label{fig:ASBL}
\end{figure}

The ASBL is an open flow that develops over a flat bottom plate in the presence of suction through that plate. 
In consequence, the BL thickness remains constant in the streamwise direction, and the ASBL shares certain properties with parallel shear flows and spatially developing BLs. 
In the DNS, the ASBL is emulated by a plane Couette setup using a high simulation domain. 
That is, we consider a fluid located in a wide gap between two parallel plates as shown schematically in Fig.~\ref{fig:ASBL}. 
The bottom plate is stationary and the fluid is set in motion through the top plate moving in the $x$-direction with velocity $U_\infty$. The latter corresponds to the free-stream velocity of the open flow. 
The flow is assumed to be incompressible and the conditions isothermal such that the density can be regarded as constant. 

\red{The occurrence of large-scale persistent coherent flow structures of long streamwise extent is one of the striking features in turbulent BLs, and ASBL is no exception.
We attempt to describe the dynamics of such a large-scale structure in a long time series using a small number of dynamic modes. 
In order to alleviate the computational effort, the simulations were carried out at moderate Reynolds number using 
a short computational domain in the streamwise direction and the sampled flow fields were averaged in streamwise direction.  
As such, the analysed two-dimensional fields obtained by streamwise averaging adequately represent the three-dimensional fields 
at least concerning the large-scale dynamics with streamwise coherence that is of interest here.
}

\subsubsection{Governing equations \& numerical details }
Expressed in units of the free-stream velocity, the laminar flow is given by
\beq
	\bm{U}=
     \begin{pmatrix}
             1-e^{-yVs/\nu} \\  
             -V_s/U_\infty \\  
             0 
     \end{pmatrix} \ , 
\eeq
where $V_s$ is the suction velocity and $\nu$ is the kinematic viscosity. The deviations $\bm{u}$ of the laminar flow are then described by the dimensionless equations
\begin{align}
	\label{eq:asbl-nse}
	\partial_t \bm{u}+\bm{u}\cdot\nabla\bm{u} + \bm{U} \cdot \nabla\bm{u} + \bm{u} \cdot \nabla\bm{U} + \nabla p -\Rey^{-1}\Delta\bm{u} = 0  \ , \nabla\cdot\bm{u} = 0 \ , 
\end{align}
where $p$ is the pressure divided by the constant density $\rho$ and $\Rey = U_\infty\delta/\nu$ the Reynolds number based on the free-stream velocity, the laminar displacement thickness $\delta = \nu/V_s$ and the kinematic viscosity $\nu$ of the fluid.

The DNS data was generated with the open-source code channelflow2.0 \cite{Gibson2014,chflow18}.
Equations \eqref{eq:asbl-nse} are solved numerically in a rectangular domain 
$\Omega = [-L_x/2,L_x/2] \times [0,H] \times [-L_z/2,L_z/2]$ as schematically shown in Fig.~\ref{fig:ASBL}, 
with periodic boundary conditions in the streamwise $x$- and the spanwise $z$-directions and no-slip 
boundary conditions in the wall-normal 
$y$-direction. Channelflow2.0 uses the standard pseudospectral technique with $2/3^{\text{rd}}$
dealiasing in stream- and spanwise directions, where the spatial discretisation
is by Fourier expansions in the homogeneous directions and a Chebyshev
expansion in the $y$-direction. The temporal discretisation is given by a third-order semi-implicit
backward differentiation scheme (SBDF3). Details of the DNS dataset are summarised in table \ref{table2}. 

\begin{table}
    \centering
    \begin{tabular}{p{0.9cm}p{0.9cm}p{1.05cm}p{0.9cm}p{0.9cm}p{0.9cm}p{0.9cm}p{0.9cm}p{0.9cm}p{0.9cm}p{0.9cm}p{1.6cm}p{0.9cm}}
        \hline
	    Re    & $\Rey_\tau$ & $\tau_w/{\rho U^2_\infty}$ & $L_x/\delta$ & $H/\delta$ & $L_z/\delta$  & $N_x$ & $N_y$ & $N_z$ & \red{$\Delta x^+ $} & \red{$ \Delta z^+ $} & $\Delta t/(\delta/U_\infty)$ & $N$ \\
        \hline
	    1000  & 320         & 0.0003                     & $4\pi$       & 20         & $4.6\pi$      & 64    & 161   & 96    & 5.1           & 3.9            &  20    & 203            \\

        \hline       
    \end{tabular}
	\caption{
	Details of the ASBL simulations discussed in Sec.~\ref{sec:asbl}.
	The Reynolds number based on the free-stream velocity $U_\infty$ and
	the laminar displacement thickness $\delta$ is denoted by $\Rey$, 
	\red{$\Rey_\tau = u_\tau \delta_{0.99}/{\nu}$ is the friction Reynolds number, 
	with $u_\tau = \sqrt{\tau_w/\rho}$ and}  
	$\tau_w$ being the shear stress at the bottom wall, $\rho$ the density, 
	\red{$\nu$ the kinematic viscosity},
        \red{$\delta_{0.99} \approx 18.5\delta$ the boundary layer thickness},  
	$L_x, H$ and $L_z$ are the length, height and width of
	the simulation domain, $N_x,~N_y$ and $N_z$ the number of grid points
	in $x$, $y$ and $z$-directions, respectively,
	\red{$\Delta x^+$ and $\Delta z^+$ the grid resolution in wall units taking into account 2/3rds dealiasing in stream- and spanwise directions,}
	$\Delta t$ the sampling interval and $N$ the number of samples. 
}
    \label{table2}
\end{table}

\subsubsection{Streaming DMD}

\begin{figure}
	\flushleft{
		\hspace{.8cm} $(a)$ \hspace{5.5cm} $(b)$ \hspace{5.2cm} $(c)$ \\
        }  
 \centering
 	\includegraphics[width=0.32\columnwidth]{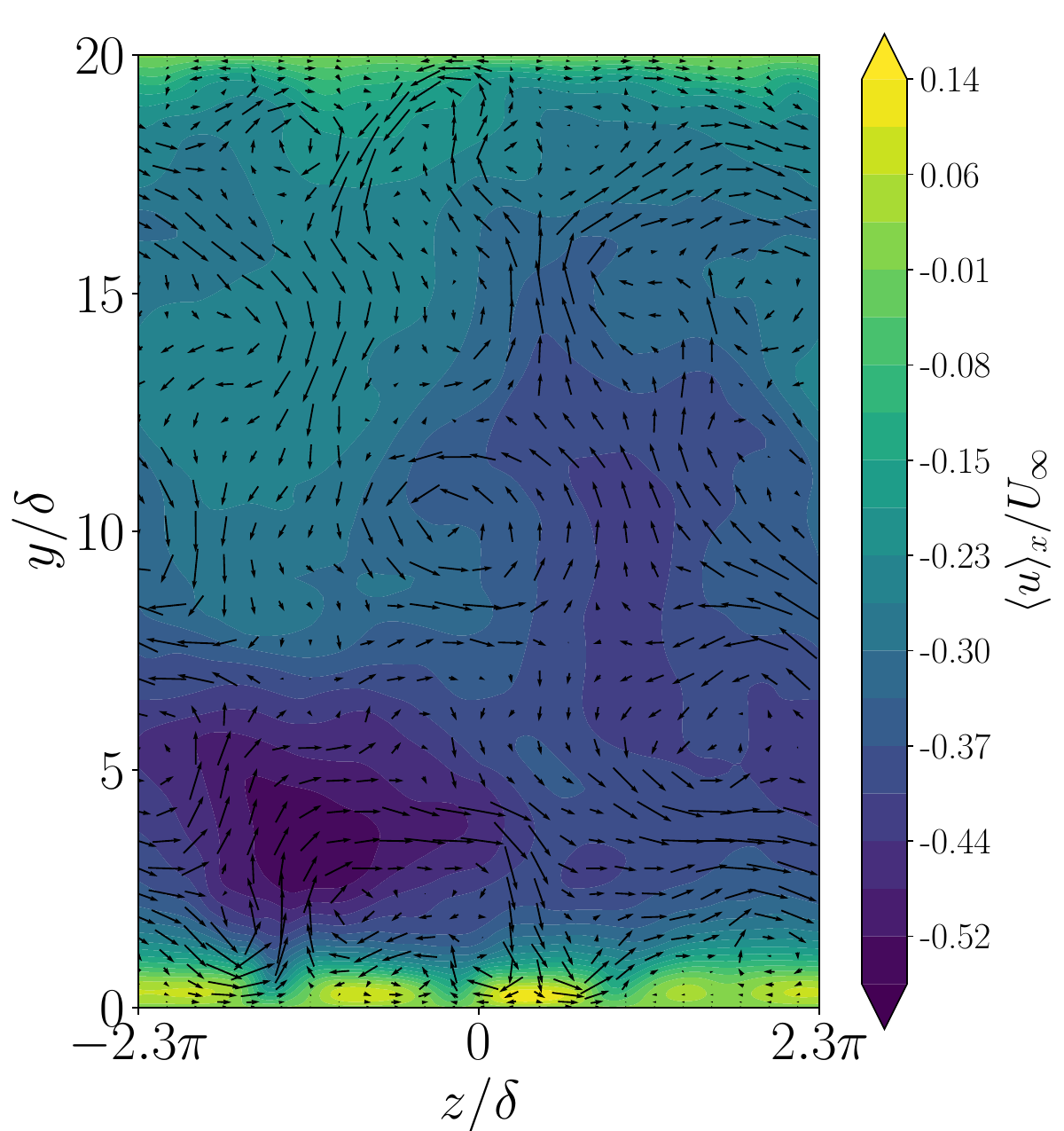}
	\includegraphics[width=0.32\columnwidth]{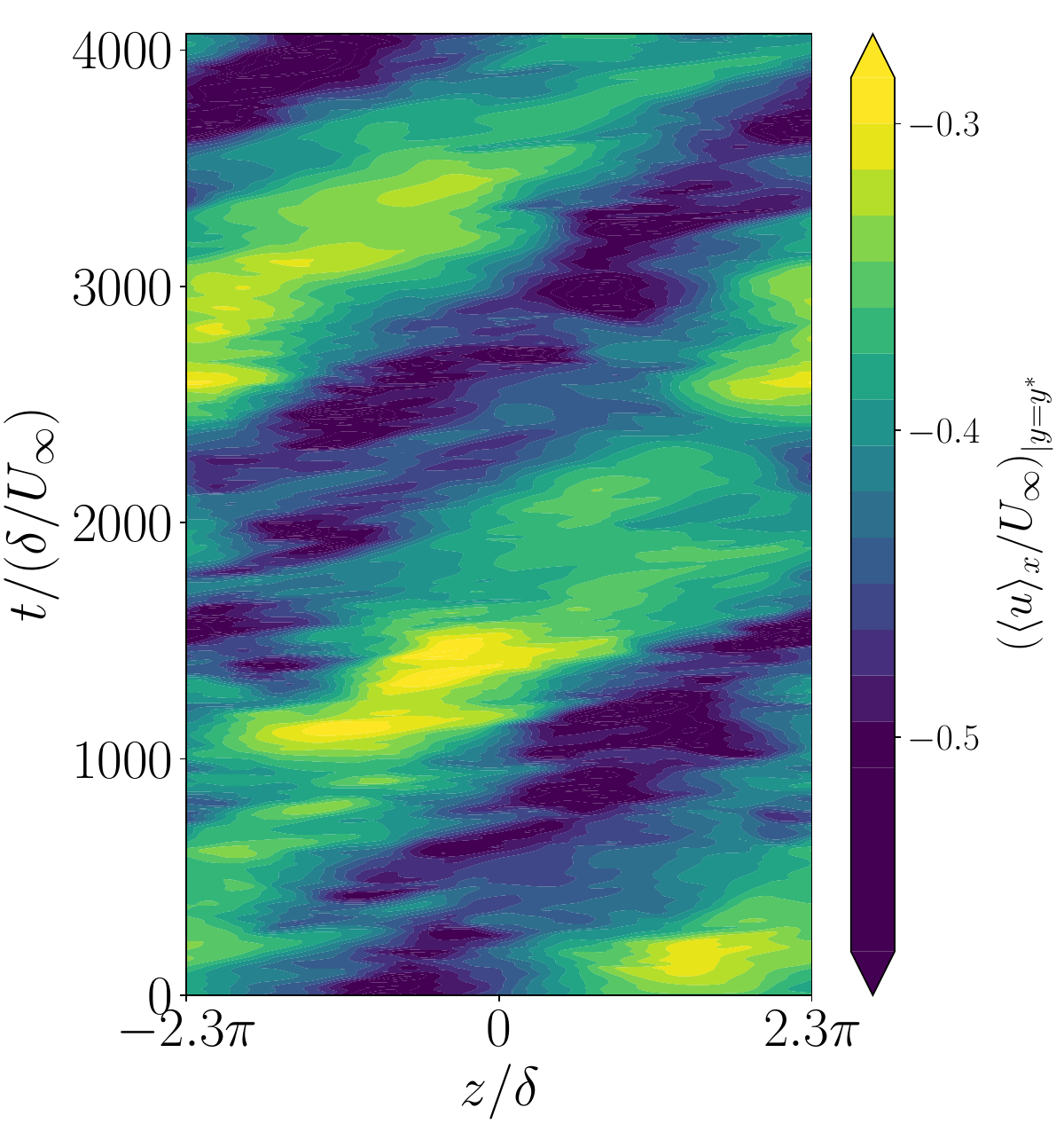}
	\includegraphics[width=0.32\columnwidth]{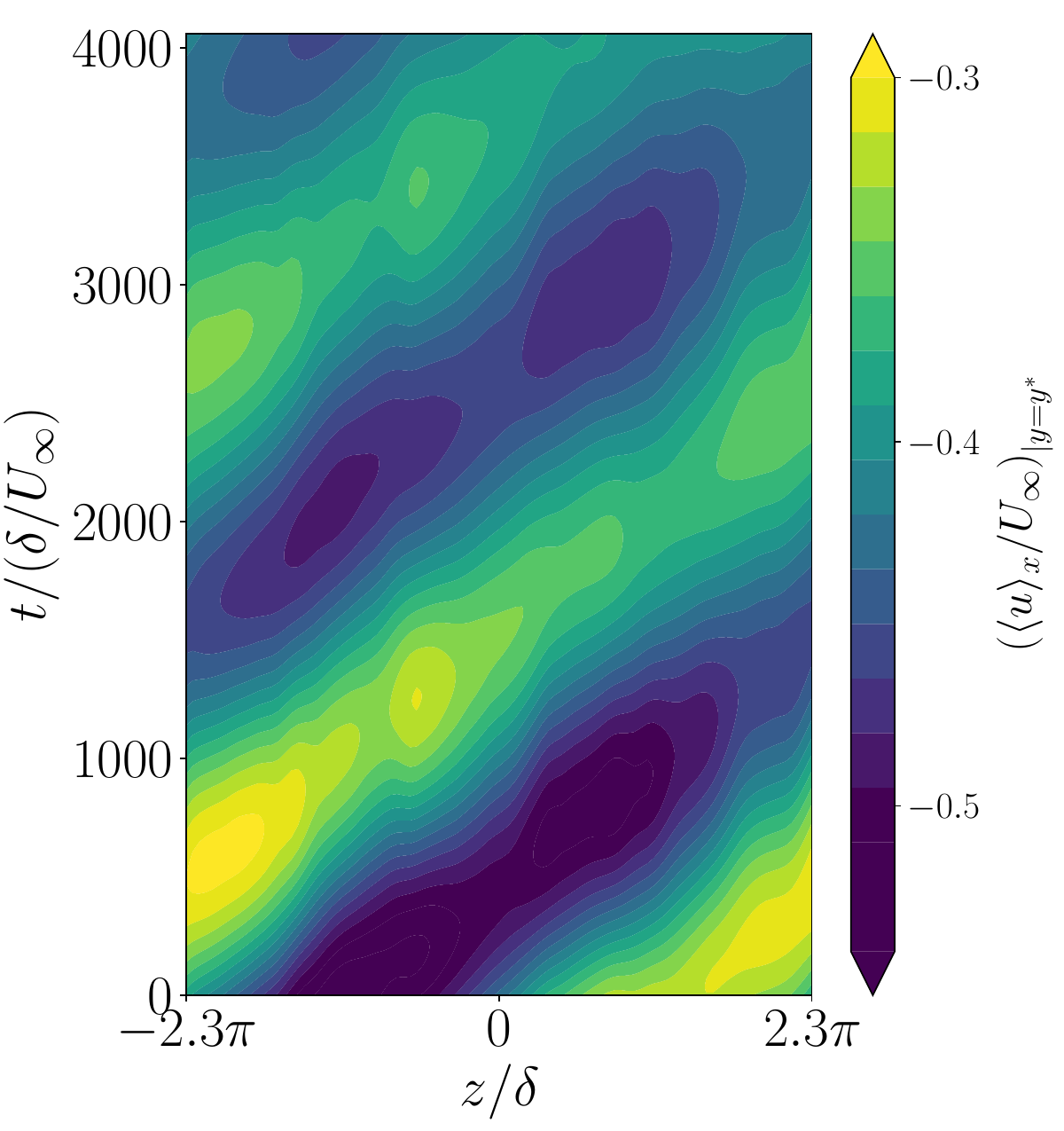}
 \caption{
	 Spatio-temporal structure and reconstruction of large-scale dynamics in the asymptotic suction boundary layer. 
	 (a) Representative original velocity-field sample \red{taken at $t = 3920 \delta/U_\infty$}. The colour coding indicates the 
	 streamwise-averaged deviation $\langle u \rangle_x$ from laminar flow in streamwise direction \red{and the arrows the streamwise-averaged cross-flow}.  
	 A slow large-scale coherent structure is clearly visible.
	 (b) Time evolution of the original flow at the centre of the coherent structure at $y^*\approx 3\delta$. 
	 (c) Time evolution of the reconstructed flow from the mean flow and the lowest frequency, that is, the first two 
	  dynamic modes at $y^*\approx 3\delta$.
 }
 \label{fig9}
\end{figure}

Figure~\ref{fig9}(a) shows the deviations from the laminar flow averaged in the streamwise direction of a typical 
data sample. A large-scale coherent region that is localised in the spanwise direction and extends from about $2\delta$ to $7\delta$ 
in wall-normal direction is clearly visible. This structure moves slower than the laminar flow and is 
accompanied by near-wall small-scale regions where the flow is faster than the laminar flow. 
The slow large-scale structure drifts through the simulation domain in spanwise direction. 
\red{
It takes the large-scale coherent structure $T = (2070\delta/U_\infty)$ time
units to traverse the simulation area once. The shift time scale $T$ and its
relative error of about $6\%$ have been determined by minimising the
$L_2$-distance between two velocity-field samples at time $t$ and $t + T'$ over
$T'$. This was repeated for several pairs of data snapshots separated by $T'$.
The time scale of the spanwise shift is also clearly discernible through 
}
the periodic pattern in the spatio-temporal evolution of the flow at a fixed
distance $y/\delta = 3$ from the bottom plate shown in Fig.~\ref{fig9}(b).
During that time it varies in intensity, as can be seen when considering the
diagonal structure visible in the spatio-temporal evolution, 
it does not disappear completely and it is difficult to discern other patterns in its dynamics. 

The aim is to reconstruct the large spatio-temporal scales of the dynamics,
i.e. the spatial extent of the slow large-scale structure, \red{its} slow
spanwise drift \red{and its dynamics},  with a few dynamic modes. Capturing the
latter \red{two} requires a very long time series and as such the application
of streaming DMD as opposed to classical DMD, as not all data can be stored in
memory at the same time.  
\red{
Following a convergence study for the first few lowest frequencies, the
truncation number was set to $r = 150$, and the sampling interval to
$20\delta/U_\infty$. This sampling interval results in about 200
velocity-field snapshots to be analysed (see table \ref{table2}). 
The lowest nonzero DMD-frequency
obtained \blue{from Eq.~\eqref{eq:eigenvalues}} results in a time scale $ T_{\rm sDMD} = 2156 \delta/U_\infty$,
which falls within the error margins of the time scale obtained by the
aformentioned minimisation procedure, $T = (2070\delta/U_\infty)$.  
}
\red{We find that three dynamic modes, that is, the mean flow and the two complex conjugate modes corresponding to the lowest frequency}, reproduce the
slow spanwise drift with \red{time scale $T_{\rm sDMD} \approx T$, here} demonstrated by comparison of the spatio-temporal evolution of the
reconstructed flow shown in Fig.~\ref{fig9}(c) with that of the original data
shown in Fig.~\ref{fig9}(b).  

\subsubsection{Streaming DMD in a co-moving frame}

The detected spanwise drift, however, is not dynamically relevant as it is merely a 
continuous shift symmetry allowed by the periodic boundary conditions in spanwise direction. In fact, DMD is known to 
perform poorly in presence of continuous symmetries \cite{kutz2016dynamic}, 
the drift can lead to spurious modes \cite{sesterhenn2019convecting}, for instance. 
Therefore, and in order to obtain further information on the 
large-scale dynamics of the ASBL, the spanwise drift was removed from the data sequence by spatially shifting each data 
sample in the sequence an appropriate distance in spanwise direction. 
That is, a re-analysis of the data was carried out in a co-moving reference frame, 
similar to the approach taken by Rowley and Marsden \cite{rowley2000symmetry}. Here, the constant shift velocity has been determined by the aforementioned 
$L_2$-minimisation, while Rowley and Marsden used a reconstruction equation to calculate a time-dependent shift velocity. 
Very recently, by combination of the method of slices for symmetry reduction \cite{budanur2015symmetry} with DMD, a new approach to 
remove time-dependent continuous symmetries has been devised \cite{marensi2021DMD}.

\begin{figure}
	\flushleft{
		\hspace{.5cm} $(a)$ \hspace{4cm} $(b)$ \hspace{3.9cm} $(c)$ \hspace{3.8cm} $(d)$ \\
        }  
	\centering
	\includegraphics[width=0.24\columnwidth]{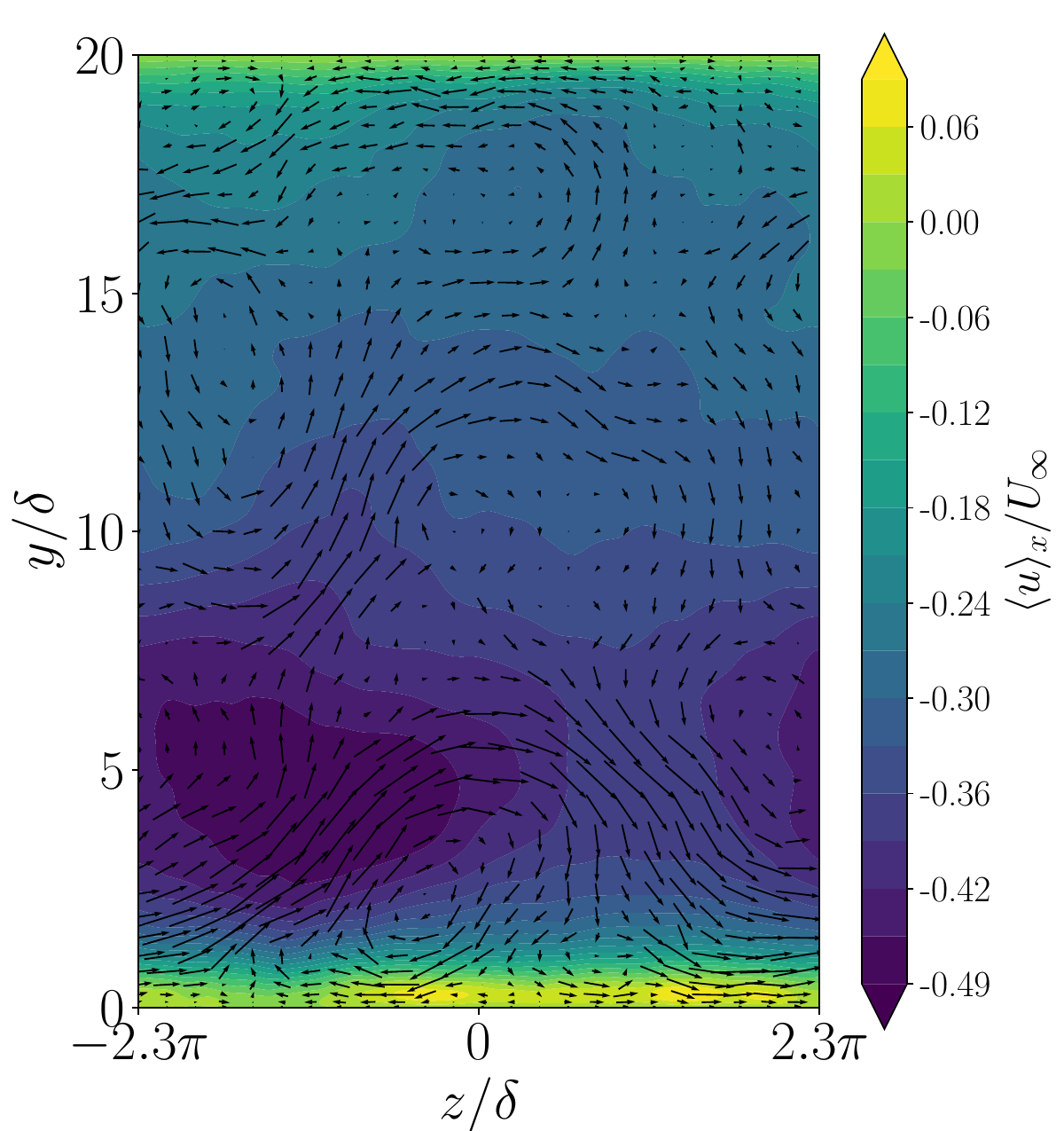}
	\includegraphics[width=0.24\columnwidth]{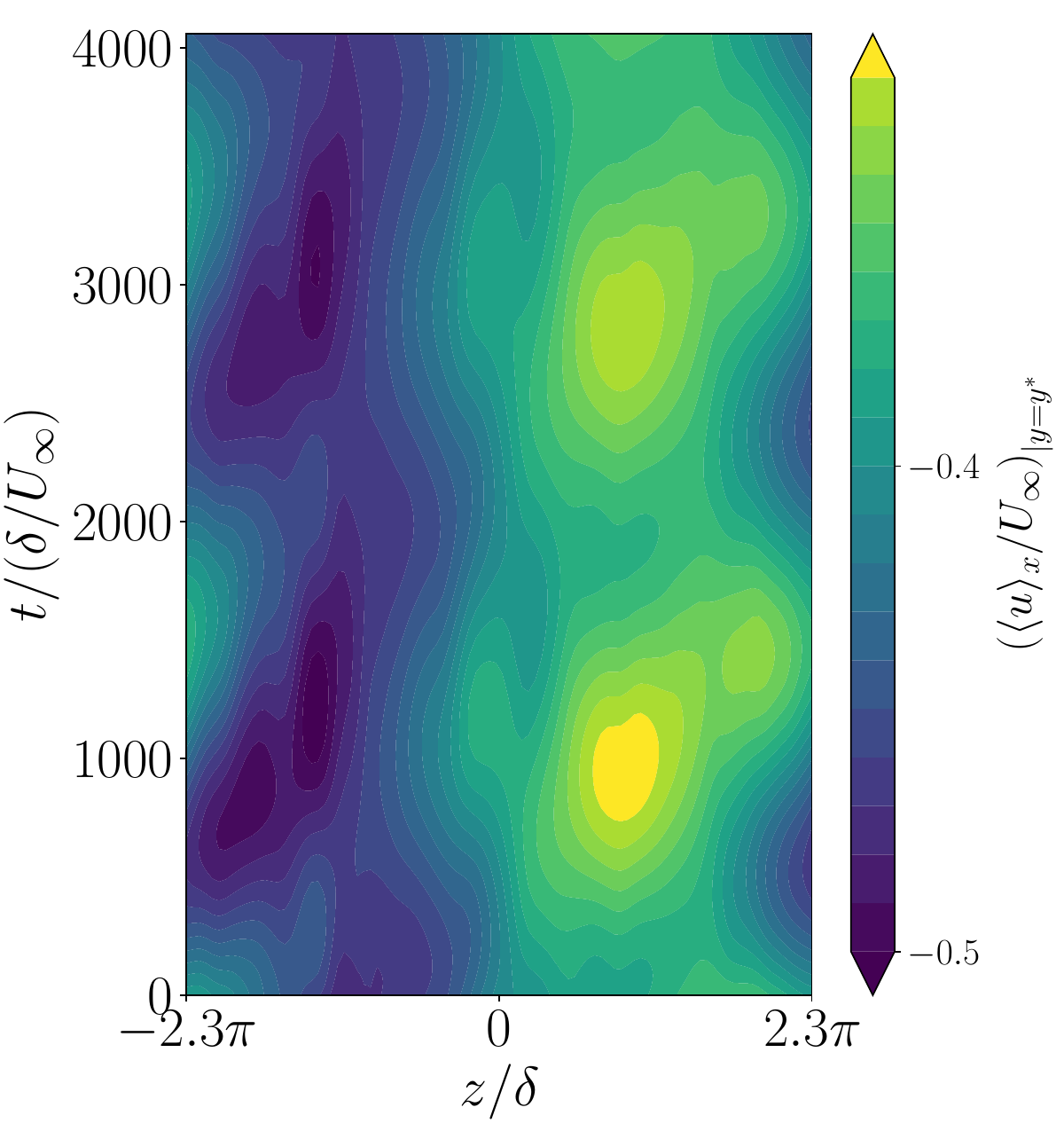}
	\includegraphics[width=0.24\columnwidth]{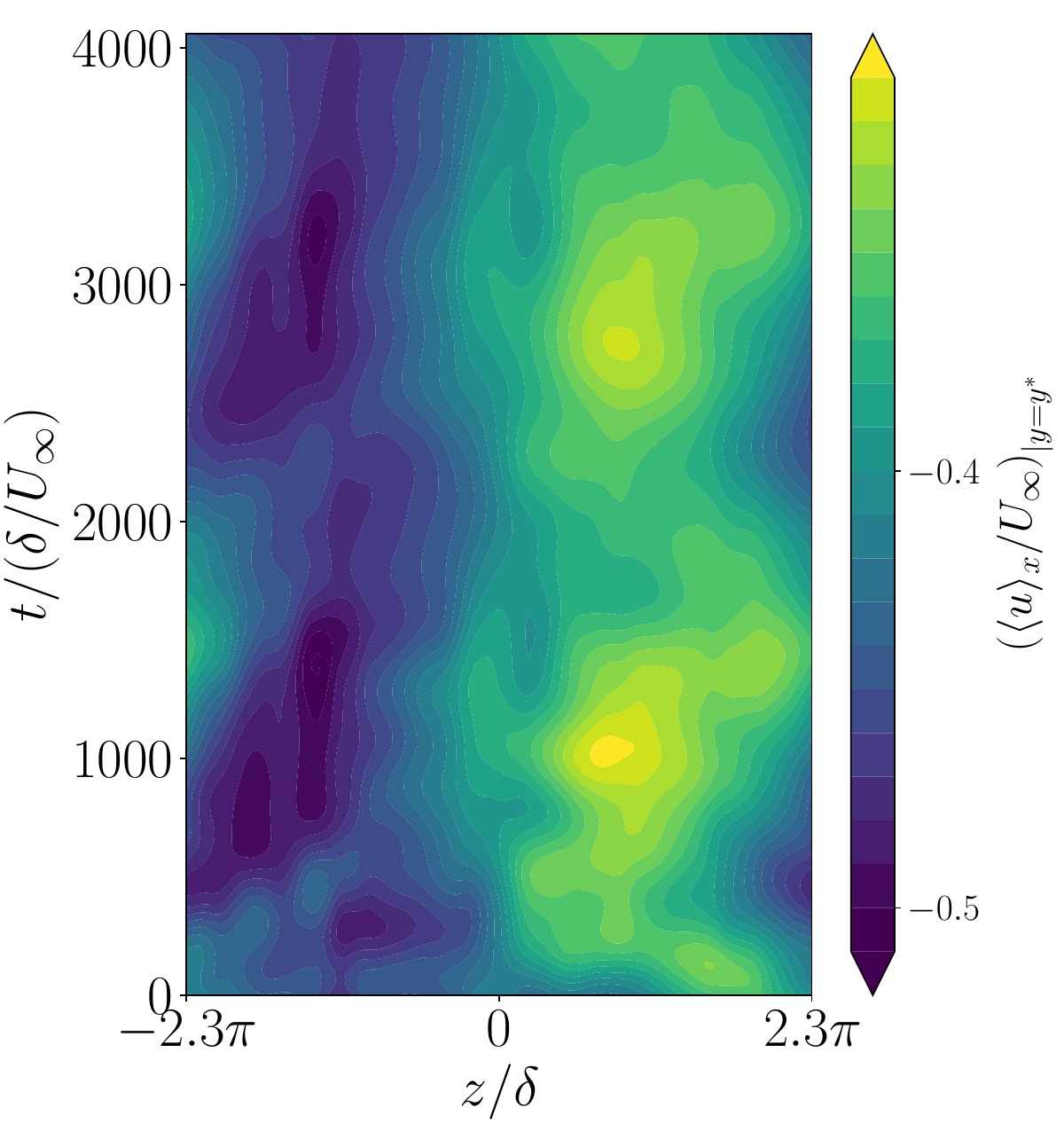}
	\includegraphics[width=0.24\columnwidth]{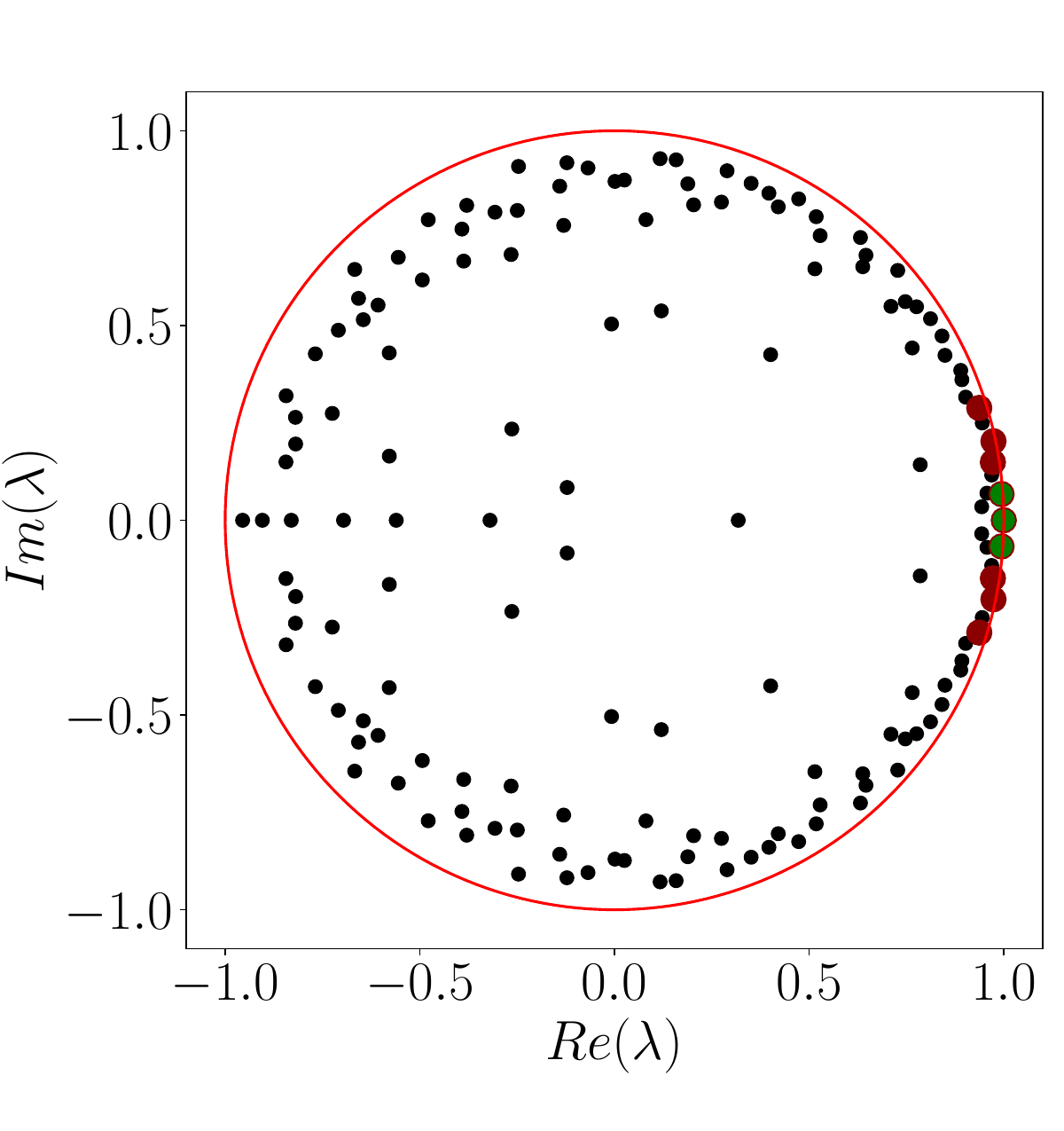}
 \caption{
         \red{Spectrum, spatio-temporal structure and reconstruction of large-scale dynamics in the asymptotic suction boundary layer
	 in a frame co-moving with the large-scale structure.
	 (a) Reconstruction of the flow at $t = 3920 \delta/U_\infty$, using two dynamic modes. 
	 Time evolution of the reconstructed flow from the first (b) two and (c) five 
	  dynamic modes at the center of the large-scale structure at $y^*\approx 3\delta$, which is located in the co-moving frame at 
	  $z/\delta \approx 1.5\pi$.
	 (d) Spectrum. The eigenvalues corresponding to the modes used in the reconstruction are shown in green (two) and red (five).
	 }
 }
 \label{fig10}
\end{figure}

In order to analyse the large-scale dynamics of the flow, we focus on flow reconstruction using low-frequency modes. 
Figure~\ref{fig10}(a) 
presents a reconstruction of the flow at \red{$t = 3920 \delta/U_\infty$, the time at which the full data sample shown in Fig.~\ref{fig9}(a) was taken,} 
using the mean flow and the complex conjugate pair of dynamic modes with the lowest frequency. 
The DMD spectrum resulting from the calculation in the co-moving frame is provided in Fig.~\ref{fig10}(d), where the eigenvalues
corresponding to the modes used in the aforementioned reconstruction are highlighted in green.
As can be seen from the data reconstruction in Fig.~\ref{fig10}(a) three dynamic modes are sufficient to reproduce the large-scale coherent structure.
However, this reconstruction does not reproduce the flow in the neighbouring regions of the coherent structure very well. For instance, 
a secondary low-momentum zone that extends further into the free stream and the vortex pattern of the original cross-flow (see Fig.~\ref{fig9}(a)) are not captured. 
We found that adding more modes, up to the highest frequency with corresponding eigenvalues still on the unit circle as indicated in red and blue in Fig.~\ref{fig10}(d), 
has very little effect (not shown).
A more adequate reconstruction could only be achieved at $r = 200$, where five dynamic modes were sufficient to describe the aforementioned
features (not shown). However, as the dataset only comprises of $203$ snapshots, $r = 200$ is likely to result in overfitting.   

The distinctive maxima and minima that are present in the spatio-temporal pattern of the 
flow evolution shown in Fig.~\ref{fig9}(b) suggest the presence of slow periodicity in the background and 
faster dynamics within the structure's core located at $z/\delta \approx 1.5 \pi$ in the co-moving frame. 
We find that the former can already be captured with the first two dynamic modes as can be seen form the corresponding spatiotemporal representation 
of the flow shown in \blue{Fig.~\ref{fig9}}(c). Adding two pairs of complex conjugate 
modes - those with eigenvalues shown in red in Fig.~\ref{fig10} \blue{(d)} - results in a representation of the background flow that changes little when reconstructed with higher frequency modes, 
see Fig.~\ref{fig10} \blue{(b)} for a spatiotemporal representation of the flow reconstructed with five modes.
The faster dynamics of the structure's core requires a higher-dimensional description. In this context we recapitulate that the structure 
never completely disappears, hence the fast dynamics encoded in the higher-frequency modes represent fluctuations in intensity on top of 
a persistent flow feature. 

\red{}

\section{Conclusion \& Outlook}
\label{sec:conclusions}

In this paper we demonstrated the applicability of streaming DMD
\cite{Hemati2014}, an efficient low-storage version of the classical SVD-based
DMD \cite{Schmid2010}, for the analysis of turbulent flows that show a certain
degree of spatio-temporal coherence.  \red{As turbulent flow dataset can be substantial in size, 
we propose a to couple the DMD calculation with prior downsampling, which is appropriate 
only if the focus is on the large-scale spatiotemporal dynamics, which we focussed on here.} 
We first validate the proposed \red{combination of downsampling with} streaming
DMD by comparing it to the classical SVD-based DMD \cite{Schmid2010}, based on
the example of the flow past a cylinder for $Re=100$. The comparison shows that
the obtained streaming dynamic modes and eigenvalues match well with those
computed from a post-processing implementation of the SVD-based DMD given
enough truncation modes.  However, streaming DMD can handle considerably larger
datasets with less computational costs compared to the SVD-based DMD, thanks to
the feature of incremental data updating, which only requires two data samples
to be held in memory at a given time. 

The objective of this study was to extract the main dynamic features with an
efficient data-driven method and use the resulting information for a
low-dimensional reconstruction of the flow. We considered three examples,
namely rapidly rotating turbulent Rayleigh--B\'enard convection, horizontal
convection, and asymptotic suction BL.  For rapidly rotating turbulent RBC, a
dominant zonal flow pattern, the boundary zonal flow, was identified through
the first two dynamic modes. Similarly, for horizontal convection two processes
that operate on different time scales could be clearly classified in terms of
dynamic modes: The second dynamic mode captures the slow oscillatory dynamics
in the bulk while the third dynamic mode describes the much faster process of
thermal plume emission. Finally, for ASBL a distinctive coherent low-momentum
zone that travels through the simulation domain in spanwise direction can be
well described by the only first two dynamic modes \red{once a streamwise drift 
is removed from the data by calculating dynamic modes in a co-moving frame of reference.
To describe dynamical features of the coherent low-momentum zone, a few more frequencies 
must be included.} These examples show that
the incrementally updated DMD algorithm can successfully decompose the dominant
structures with corresponding frequencies and modes. This establishes sDMD as
an accurate and efficient method to identify and capture dominant
spatio-temporal features from large datasets of highly-turbulent flows.

As DMD decomposes datasets into coherent structures based on characteristic
frequencies, it is especially useful for the analysis of flows featuring
large-scale coherent structures and periodic motion. The advantages of the sDMD
algorithm, both in terms of low-storage and potential real-time implementation,
will make DMD available in numerous contexts where it would have been
infeasible previously. This includes in particular the analysis of massive
datasets that cannot completely reside in memory. One such application, for
instance, concerns the search for unstable periodic orbits in turbulent flows,
where a classical DMD-based approach has been successfully applied at moderate
Reynolds number \cite{Page2020}. Streaming DMD may constitute a step forward in
extending the applicability of this method to higher Reynolds numbers. 

In further steps, streaming DMD can be applied to different turbulent
flow datasets, to investigate in detail the ability to decompose coherent flow
structures. One disadvantage of streaming DMD is that the truncation number of
snapshots to achieve a similar reconstruction accuracy as the SVD-based DMD
is typically larger.  Here, \red{as we focussed on statistically stationary large spatio-temporal scales only,} 
apart from convergence tests we have not considered the effect of the truncation
number but have \red{restricted our attention} on the analysis of the first few 
\red{modes corresponding to statistically stationary dynamics ranked by frequency in ascending order.} 
The truncation number is, however, important and should be
quantitatively considered when applying streaming DMD to flow field
reconstruction or decomposition of complex turbulent flows with multi-frequency
temporal structures. \red{Similarly, the mode ordering by frequency used here may not be the 
optimal choice for all datasets.  An efficient and robust criterion needs to be introduced for rank selection.}

Finally we wish to mention that not only DMD but also some other approaches,
based on or related to DMD, might be very efficient in extraction and analysis
of the dynamics of the turbulent superstructures.  While in the DMD we apply
linear transformations to obtain modes out of snapshots and vice versa, a
natural extension of the DMD would be to employ non-linear transformations
instead. This can be realized either via application of hand-picked nonlinear
functions (e.g. to use so-called extended DMD -- eDMD \cite{Williams2015}),
\red{by calculation of the full Koopman modes (see \cite{giannakis2018koopman} in the context of convection)},
or by training a deep neural network (e.g. to use deep
Koopman models \cite{Morton2018DeepDM}). A  multilayer convolutional
neural network appears to be a good candidate for such a task.  In general,
(un)supervised deep learning seems to be very promising \red{technique for} 
extraction and analysis of the global dynamics of the turbulent flow
superstructures. 
\blue{Further to this, kernel
methods or tensor-based reformulations of DMD exist that are also well suited for high-dimensional data sets.}
A more detailed consideration of these alternate approaches
is beyond the scope of this article and application of these advanced methods
for the turbulent superstructure analysis remains a challenge for future
studies.

\subsection*{Acknowledgements}
This work is supported by the Max Planck Center for Complex Fluid Dynamics, the Priority Programme SPP 1881 ``Turbulent Superstructures" of the Deutsche
Forschungsgemeinschaft (DFG) under grants Sh405/7 and Li3694/1 and DFG grants Sh405/8 and Sh405/10. The authors acknowledge the Leibniz Supercomputing Centre (LRZ) and
the Lichtenberg high performance computer of the TU Darmstadt for providing computing time. \red{This work used the ARCHER UK National Supercomputing Service 
({\tt http://www.archer.ac.uk}).} 


\end{document}